\documentclass[12pt, draftcls, onecolumn]{IEEEtran}

\usepackage{bm}
\usepackage{verbatim}
\usepackage{algorithm}
\usepackage{soul,xcolor}
\usepackage[T1]{fontenc}
\usepackage{algcompatible}
\usepackage[normalem]{ulem}
\usepackage{multirow,enumitem}
\usepackage{graphicx}

\usepackage{url}
\usepackage{bbm}
\usepackage{cite}
\usepackage{array}
\usepackage{ifthen}
\usepackage{xspace}
\usepackage{dsfont}
\usepackage{amsmath}
\usepackage{amssymb}
\usepackage{multicol}
\usepackage{amsfonts}
\usepackage{mathrsfs}
\usepackage{booktabs}
\usepackage{caption}
\usepackage{subcaption}
\usepackage{algpseudocode}

\usepackage{esint}
\usepackage{amsthm}
\usepackage{cancel}
\usepackage{siunitx}
\usepackage{setspace}
\usepackage{hyperref}
\usepackage{makecell}
\usepackage{footnote}

\theoremstyle{plain}

\theoremstyle{definition}

\theoremstyle{remark}

\newtheorem{prop}{\textbf{Proposition}}

\DeclareMathOperator*{\argmax}{arg\,max}
\DeclareMathOperator*{\argmin}{arg\,min}

\newcommand\hlt[1]{\textcolor{black}{#1}}

\title{MAESTRO-X: Distributed Orchestration of Rotary-Wing UAV-Relay Swarms}
\author{Bharath Keshavamurthy\IEEEauthorrefmark{1}, Matthew A. Bliss\IEEEauthorrefmark{2}, and Nicol\`{o} Michelusi\IEEEauthorrefmark{1}
\thanks{A preliminary version of this work was presented at Asilomar 2022 \cite{ASILOMAR}. Source code is available on \href{https://github.com/bharathkeshavamurthy/MAESTRO-X.git}{GitHub} \cite{MAESTRO-X}.}
\thanks{Part of this work has been supported by NSF under grants CNS-1642982 and CNS-2129015.}
\thanks{\IEEEauthorrefmark{1}Electrical, Computer and Energy Engineering, Arizona State University, Tempe, AZ.}
\thanks{\IEEEauthorrefmark{2}Electrical and Computer Engineering, Purdue University, West Lafayette, IN.}
\vspace{-6mm}
}

\begin{document}
\bstctlcite{IEEEexample:BSTcontrol}

\maketitle
\thispagestyle{plain}
\pagestyle{plain}
\setulcolor{red}
\setul{red}{2pt}
\setstcolor{red}
\vspace{-12mm}
\begin{abstract}
This work details a scalable framework to orchestrate a swarm of rotary-wing UAVs serving as cellular relays to facilitate beyond line-of-sight connectivity and traffic offloading for ground users. First, a Multiscale Adaptive Energy-conscious Scheduling and TRajectory Optimization (MAESTRO) framework is developed for a single UAV. Aiming to minimize the time-averaged latency to serve user requests, subject to an average UAV power constraint, it is shown that the optimization problem can be cast as a semi-Markov decision process, and exhibits a multiscale structure: outer actions on radial wait velocities and terminal service positions minimize the long-term delay-power trade-off, optimized via value iteration; given these outer actions, inner actions on angular wait velocities and service trajectories minimize a short-term delay-energy cost; finally, rate adaptation is embedded along the trajectory to leverage air-to-ground channel propagation conditions. A novel hierarchical competitive swarm optimization scheme is developed in the inner optimization, to devise high-resolution trajectories via iterative pair-wise updates. Next, MAESTRO is eXtended to UAV swarms (MAESTRO-X) via scalable policy replication, enabled by a decentralized command-and-control network augmented with: 
(1) \emph{spread maximization} to proactively position UAVs to serve future requests;
(2) \emph{consensus-driven conflict resolution} to orchestrate scheduling decisions based on delay-energy costs including queuing dynamics;
(3) \hlt{\emph{adaptive frequency reuse} to improve spectrum utilization across the network}; and 
(4) \hlt{a \emph{piggybacking mechanism} allowing UAVs to serve multiple ground users simultaneously}. 
\hlt{Numerical evaluations show that, for user requests of 10 Mbits, generated according to a Poisson arrival process with rate 0.2 req/min/UAV, single-agent MAESTRO offers 3.8$\times$ faster service than a high-altitude platform and 29\% faster than a static UAV deployment; moreover, for a swarm of 3 UAV-relays, MAESTRO-X delivers data payloads 4.7$\times$ faster than a successive convex approximation scheme; and remarkably, a single UAV optimized via MAESTRO outclasses 3 UAVs optimized via a deep-Q network by $38$\%.}
\end{abstract}
\vspace{-4mm}

\vspace{-4mm}

\section{Introduction}\label{S1}
Enterprises across various industrial sectors have stepped-up the adoption of Unmanned Aerial Vehicles (UAVs) to gather data, survey infrastructure, monitor operations, and automate logistics \cite{UAVSurvey, UAVTutorial}.
UAVs can also be leveraged to enhance troop deployments in military scenarios \cite{TCCN}, aid emergency response during a natural disaster \cite{VerizonDisasterRelief}, and facilitate data harvesting in precision agriculture \cite{VerizonAgriculture}. Inevitably, this has fostered varied academic research and industrial R\&D on UAV-augmented beyond line-of-sight connectivity and traffic offloading in cellular networks, whose coverage can be enhanced by the mobility and maneuverability of UAVs \cite{LOSDominance, FundamentalTradeoffs}. 

Yet, the pervasive potential of UAV-assisted wireless networks presents a plethora of challenges in real-world deployments \cite{FundamentalTradeoffs}: limited on-board energy of aerial platforms, Quality-of-Service (QoS) requirements, air-to-ground channels, and computational feasibility challenges of UAV trajectory design. Several works have tackled some of these challenges by employing tools from optimization and artificial intelligence---however, numerous problems remain unsolved: failure to capture uncertain system dynamics vis-\`{a}-vis random traffic arrivals \cite{SCA, PAoI, MEC-CVX, LoSMap, Rician}; restrictions on UAV path and velocity characteristics \cite{PSOPathStructure, PAoI}; inefficient centralized swarm deployments \cite{CSCA-ADMM, JointTrajectoryDesign, MultiDroneDeployment}; computationally expensive joint multi-agent formulations offering limited scalability \cite{DDQN, MEC-DDPG, DQNPositioning, MLDeployment}; and failure to account for link layer effects on the QoS of the network \cite{GameTheory, UAVDynamicCoverage}.

In this paper, considering these drawbacks in the state-of-the-art, we study the decentralized orchestration of multiple power-constrained rotary-wing UAVs supplementing a terrestrial base station by relaying data traffic dynamically generated by ground users. \hlt{Incorporating waiting state optimization, computationally feasible trajectory design, throughput-maximizing rate adaptation to Air-to-Ground (A2G) propagation conditions, queue management, frequency reuse to enhance spectrum utilization, multi-user service, and multi-UAV consensus-driven scheduling, we develop a scalable framework to efficiently automate the operations of distributed UAV-relay deployments.}

Ergo, specializing to single UAV-relay settings, we first propose MAESTRO, a Multiscale Adaptive Energy-conscious Scheduling and TRajectory Optimization framework to control the idle and service phase operations of the UAV. Seeking to minimize the average communication delay subject to an average UAV mobility power constraint, we show that the problem can be cast as a Semi-Markov Decision Process (SMDP) with a multiscale structure: outer decisions on radial velocities and terminal service positions influence the long-term delay-power cost; consequently, given these outer actions, inner actions on angular wait velocities and service trajectories minimize a short-term delay-energy cost. \hlt{We develop a value iteration algorithm \cite{Bertsekas} exploiting this multiscale structure to optimize outer actions, and a hierarchical variant of Competitive Swarm Optimization (CSO) \cite{CSO}, decoupled from value iteration, to optimize high-resolution trajectories embedding a novel throughput maximizing rate adaptation scheme for A2G channels. Next, we extend MAESTRO to a swarm of UAV-relays (MAESTRO-X) via a scalable replication strategy, enabled by a decentralized command-and-control network and augmented with: spread maximization to proactively position the UAVs to serve future service requests; consensus-driven conflict resolution to orchestrate ground user scheduling decisions based on delay-energy costs, including queuing dynamics; frequency reuse to enhance spectrum utilization; and piggybacking to enable each UAV to serve multiple users simultaneously.}
\begin{table}
\begin{center}
\scriptsize
    \begin{tabular}{|*{12}{c|}}
    \hline
    	\multirow{ 2}{*}{{\bf{Paper}}} &
    	{\bf{Adaptive}} &
	\bf{Channel} &
    	\bf{\hlt{Frequency}} &
   	\bf{\hlt{Multiuser}} &
	\multicolumn{2}{c|}{\bf{UAV Motion}} &
    	\bf{UAV} &
    	\bf{Multi-UAV} &
	\bf{Overall} &
	\multicolumn{2}{c|}{\bf{Link Layer}}\\
		&
    	\bf{control} &
	\bf{model} &
    	\bf{\hlt{reuse}} &
   	\bf{\hlt{service}} &
	\bf{Mobility} & \bf{Velocity} &
    	\bf{deployment} &
    	\bf{scheduling} &
	\bf{formulation} &
	\bf{Schedule} & \bf{Queue}\\
    \hline
	{\tiny\bf MAESTRO-X} & Yes & A2G & \hlt{Yes} & \hlt{Yes} & Dynamic & Variable & Distributed & Decoupled & Model-based & Yes & Yes\\
	\hline
    \cite{SCA} & No & FSPL & \hlt{No} & \hlt{No} & Dynamic & Variable & Single & - & Model-based & Yes & No\\
    \hline
    \cite{CSCA-ADMM} & No & A2G & \hlt{Yes} & \hlt{Yes} & Dynamic & Variable & Centralized & Joint & Model-based & Yes & No\\
    \hline
    \cite{DDQN} & No & A2G & \hlt{No} & \hlt{Yes} & Restricted & Fixed & Distributed & Joint & Model-free & No & No\\
    \hline
    \cite{PAoI} & No & FSPL & \hlt{No} & \hlt{No} & Dynamic & Fixed & Single & - & Model-based & Yes & No\\
    \hline
    \cite{MEC-CVX} & No & FSPL & \hlt{No} & \hlt{No} & Dynamic & Variable & Single & - & Model-based & Yes & No\\
    \hline
    \cite{MEC-DDPG} & No & FSPL & \hlt{No} & \hlt{Yes} & Restricted & Fixed & Distributed & Joint & Model-free & Yes & No\\
    \hline
    \cite{LoSMap} & No & A2G & \hlt{No} & \hlt{No} & Static & - & Single & - & Model-based & No & No\\
    \hline
    \cite{GameTheory} & No & FSPL &\hlt{No} & \hlt{No} &  Static & - & Distributed & Joint & Model-based & Yes & No\\
    \hline
    \cite{UAVDynamicCoverage} & Yes & FSPL & \hlt{No} & \hlt{No} & Static & - & Distributed & Joint & Model-based & No & No\\
    \hline
    \cite{JointTrajectoryDesign} & No & FSPL & \hlt{No} & \hlt{No} & Dynamic & Fixed & Centralized & Joint & Model-based & Yes & No\\
    \hline
    \cite{MultiDroneDeployment} & No & A2G & \hlt{No} & \hlt{No} & Static & - & Centralized & Joint & Model-based & No & No\\
    \hline
    \cite{RLSenseSend} & No & A2G & \hlt{No} & \hlt{No} & Restricted & Fixed & Distributed & Decoupled & Model-free & No & No\\
    \hline
    \cite{DQNPositioning} & Yes & FSPL & \hlt{No} & \hlt{No} & Static & - & Distributed & Joint & Model-free & No & Yes\\
    \hline
    \cite{MLDeployment} & Yes & A2G & \hlt{No} & \hlt{No} & Static & - & Distributed & Joint & Model-free & No & No\\
    \hline
    \cite{Rician} & No & A2G & \hlt{No} & \hlt{No} & Dynamic & Variable & Single & - & Model-based & Yes & No\\
    \hline
    \cite{UAV-DRL} & Yes & FSPL & \hlt{No} & \hlt{No} & Dynamic & Variable & Single & - & Model-free & No & No\\
    \hline
    \end{tabular}
    \vspace{-2mm}
    \caption{A comparison of the features of our framework with those of relevant schemes in the literature.}
    \label{T1}
\end{center}
\vspace{-4mm}
\end{table}

\noindent{\textbf{Related Work}}: Table~\ref{T1} summarizes our approach (MAESTRO-X) and contrasts it with relevant works in the state-of-the-art. First, we observe non-adaptive schemes, e.g., \cite{SCA, JointTrajectoryDesign, MultiDroneDeployment} designed for applications where ground users possess local storage or aggregation capabilities allowing for deterministic traffic; however, practical deployments involve dynamically generated requests and randomly located ground users. Accommodating these uncertainties calls for the design of adaptive UAV orchestration frameworks. Yet, existing works do so only for single UAV-relay deployments \cite{UAV-DRL} or consider static placement of UAVs (i.e., no trajectory design) \cite{UAVDynamicCoverage, DQNPositioning, MLDeployment}. In contrast, we design adaptive trajectory and scheduling strategies for distributed multi-UAV swarms, that accommodate dynamic and uncertain traffic generated by ground users.

\hlt{Next, works employing Free Space Pathloss (FSPL) channel models, e.g., \cite{SCA, PAoI, MEC-CVX, MEC-DDPG}, fail to account for the A2G channel characteristics in UAV-assisted wireless networks. 
Existing works that model A2G channels fail to leverage small- and large-scale A2G conditions via rate adaptation. A notable exception is \cite{Rician}, which differs from our rate adaptation scheme in two ways: 1) we select the rate to maximize throughput (vs. \cite{Rician}, which aims to satisfy an outage constraint), and 2) we use a probabilistic line-of-sight (LoS) and Non-LoS (NLoS) model. Furthermore, most works surveyed neither consider spectrum reuse (with the exception of \cite{CSCA-ADMM}) nor permit simultaneous multi-user service (with the exception of \cite{CSCA-ADMM, DDQN, MEC-DDPG})---however, the works that do incorporate these crucial features \cite{CSCA-ADMM, DDQN, MEC-DDPG} fail to consider adaptation to dynamically generated requests from randomly located users, as done in our work.}

A common approach for trajectory design is Successive Convex Approximation (SCA) \cite{SCA, Rician}. SCA typically relies on the FSPL channel model to devise convex relaxations of the objective and constraints. Exceptions include \cite{Rician} and \cite{CSCA-ADMM}, which apply SCA approaches under A2G channels. In \cite{Rician}, a logistic approximation of the achievable rate is used under outage constraints; in \cite{CSCA-ADMM}, only large-scale fading is considered. However, when coupling trajectory design with our throughput-maximizing rate adaptation scheme, closed-form rate expressions with first-order convex approximations are impractical. To tackle this challenge, we propose a CSO \cite{CSO} approach for UAV trajectory design. \hlt{Unlike SCA, CSO does not rely on the problem structure of FSPL models to work effectively, and can thus accommodate realistic A2G propagation conditions.} Particle Swarm Optimization (PSO) \cite{PSO}, a swarm-based optimization method in which particle updates are driven by the global and individual best positions, has been used to optimize static UAV placement \cite{Efficient3DPlacementPSO, 3DDeploymentPSO}, or restricted UAV trajectories (e.g., moving along a circle \cite{PSOPathStructure}, or with fixed speed \cite{PAoI}). Removing these restrictions calls for the more efficient update strategy of CSO, which exhibits superior performance on several benchmarks \cite{CSO}: it involves pair-wise particle competitions, wherein winners advance to the next iteration and the losers learn from the winners. \hlt{Moreover, we scale CSO to higher-dimensional trajectory design by embedding it within a Hierarchical wrapper (HCSO), which iteratively optimizes trajectories of increasing resolution, without imposing unreasonable restrictions on UAV mobility.}

Next, shifting our attention to swarm orchestration frameworks, several approaches consider centralized multi-UAV deployments \cite{JointTrajectoryDesign, MultiDroneDeployment, CSCA-ADMM} in which an aggregation center coordinates the UAV-relaying operations; or either joint multi-relay solutions \cite{CSCA-ADMM, GameTheory, UAVDynamicCoverage} or model-free formulations constituting combined state and action spaces \cite{DDQN, MEC-DDPG, DQNPositioning, MLDeployment}. \hlt{An exception is \cite{RLSenseSend}, which considers a model-free setup with decentralized UAV deployments and decoupled scheduling. But, \cite{RLSenseSend} does not consider adaptation to randomly-generated data traffic, as we do in our work; rather, a sense-and-send protocol is devised, wherein tasks are always ready to be sensed.} Centralized swarm deployments often need additional capital and operational expenditure, and joint multi-UAV designs lead to large solution spaces resulting in prohibitive convergence times. Mindful of such considerations, we present an orchestration framework suitable for distributed UAV deployments by replicating our single-agent policy across the swarm and augmenting it with spread maximization and consensus-driven link-layer prescient conflict resolution over a command-and-control network. This eliminates the need for a centralized aggregation center, mitigates the computational overhead encountered by joint multi-relay models, and facilitates the seamless incorporation of queuing dynamics into scheduling decisions. Also, as shown in our numerical evaluations, our framework can be scaled to networks with $\geq$10 UAVs, while state-of-the-art approaches \cite{SCA, CSCA-ADMM, DDQN} become prohibitively expensive for networks with 5 UAVs. Additionally, although model-free control schemes \cite{DDQN, MEC-DDPG, RLSenseSend, DQNPositioning, MLDeployment, UAV-DRL} consider unknown system dynamics when solving for the optimal trajectory and/or scheduling solution, they fail to efficiently exploit the problem structure, resulting in large policy convergence times. In contrast, we use a model-based approach, by casting the problem as an SMDP, which captures the temporal irregularities seen in the state transitions of UAV-augmented wireless networks.

\phantomsection\label{contrib_subs}
\noindent{\hlt{\textbf{Contributions}}}: \hlt{We develop a novel framework for the scalable orchestration of UAV-relay swarms. To the best of our knowledge, no other work simultaneously incorporates the practical features of 1) dynamic traffic from randomly located ground users; 2) efficient exploitation of A2G channel conditions via a throughput-maximizing rate adaptation scheme; 3) easy scalability to large UAV swarms via policy replication, coupled with multi-agent coordination mechanisms over a distributed command-and-control network; and 4) waiting state optimization to position idle UAVs for potential new requests. In a nutshell, the contributions of this paper are:
\begin{itemize}[leftmargin=*]
    \item \underline{MAESTRO}: For a single UAV, we construct an adaptive scheduling and trajectory design framework to minimize the communication latencies in serving dynamic transmission requests generated by randomly located ground users, subject to an average UAV power constraint. We show that the problem can be solved as a \emph{Semi-Markov Decision Process} (SMDP). A multiscale decomposition facilitates efficient computation of rate adaptation, scheduling and trajectory solutions, and energy-conscious orchestration of the UAV during idle periods.
    \item \underline{HCSO}: To enable computationally tractable design of high-resolution UAV trajectories under A2G propagation conditions, we propose \emph{Hierarchical CSO} (HCSO), a variant of CSO wherein iterative pair-wise cost comparisons devise trajectories of increasingly higher resolution.
    \item \underline{MAESTRO-X}: Coupled with decentralized command-and-control operations over a distributed mesh network, we augment the single-UAV trained policy with multi-UAV mechanisms to orchestrate waiting phase operations (\emph{spread maximization}), coordinate scheduling decisions incorporating queuing dynamics (\emph{consensus-driven conflict resolution}), enable simultaneous multi-user service (\emph{piggybacking}), and enhance spectrum utilization (\emph{frequency reuse}).
\end{itemize}}

The rest of the paper is organized as follows: Sec.~\ref{S2} introduces the system model; Sec.~\ref{S3} elucidates the design of MAESTRO; Sec.~\ref{S4} describes the main algorithms; Sec.~\ref{S5} details policy replication and multi-UAV mechanisms to manage distributed swarms (MAESTRO-X); Sec.~\ref{S6} chronicles our numerical evaluations; and finally, Sec.~\ref{S7} lists our concluding remarks.
\vspace{-4mm}

\section{System Model}\label{S2}
\begin{figure} [t]
    \begin{subfigure}{0.505\linewidth}
        \centering
        \includegraphics[width=1.0\linewidth]{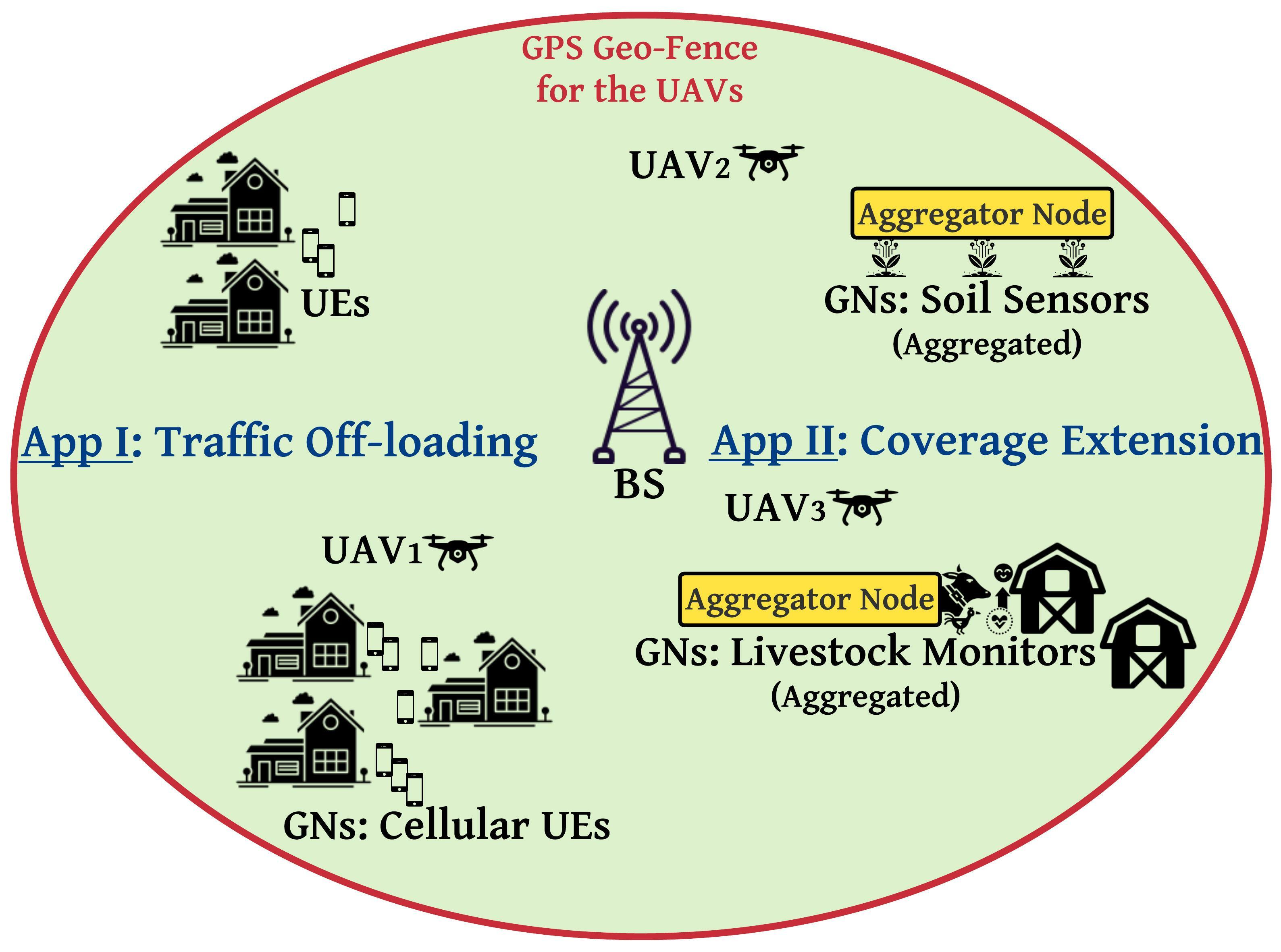}
        \caption{Deployment Model.}
        \label{F1}
    \end{subfigure}
    \hfill
    \begin{subfigure}{0.44\linewidth}
        \centering
        \includegraphics[width=1.0\linewidth]{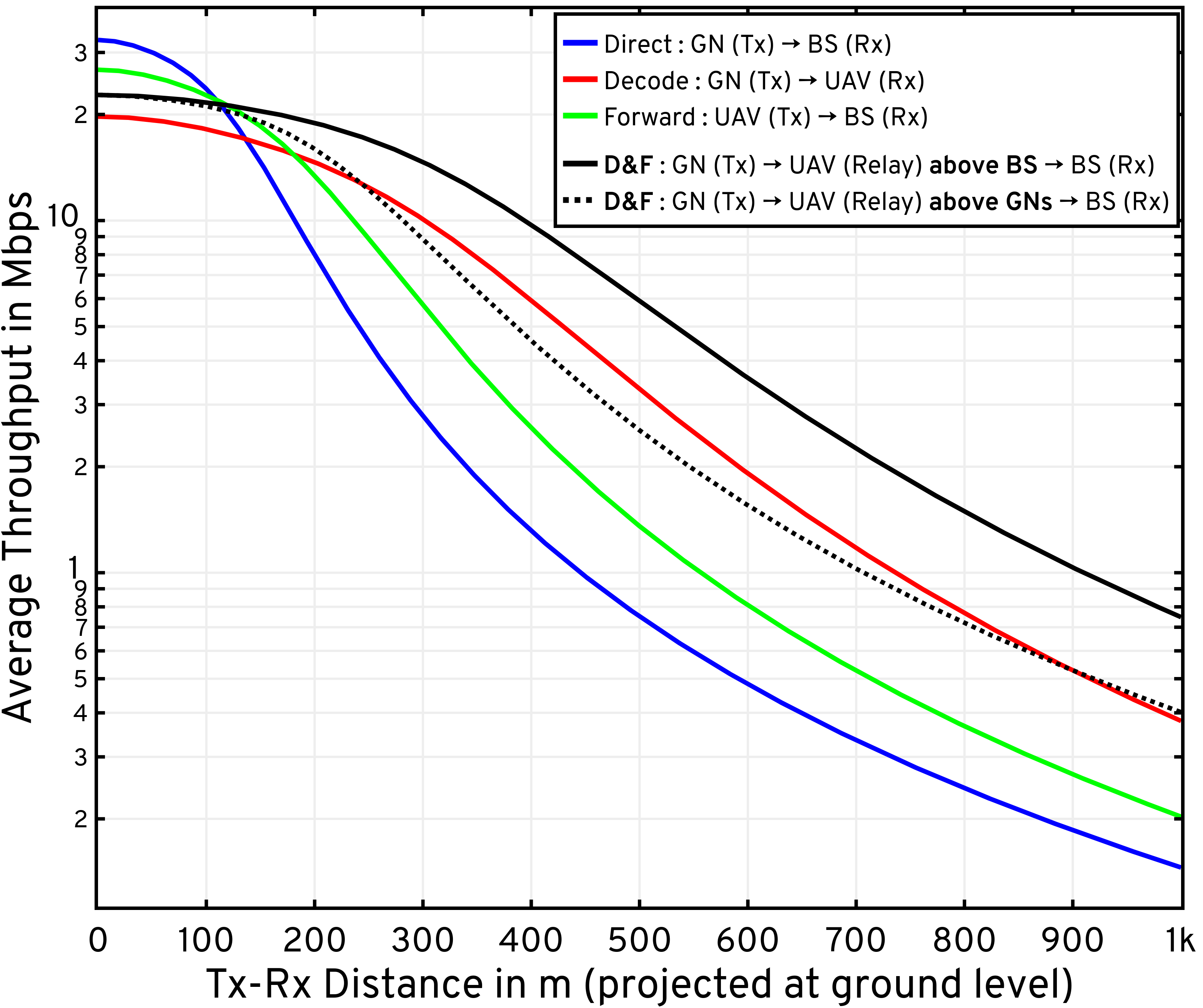}
        \caption{\hlt{Throughputs under A2G channel.}}
        \label{F2}
    \end{subfigure}
    \vspace{-2mm}
    \caption{(a) A terrestrial BS aided by UAVs serving as relays for a diverse set of GNs: traffic offloading for cellular UEs, and coverage extensions for livestock monitors and soil sensors; \hlt{(b) rate-adapted throughputs (see Table \ref{T_params} for the numerical parameters) along the GN${\rightarrow}$BS link (\emph{direct}), GN${\rightarrow}$UAV link (\emph{decode}),  UAV${\rightarrow}$BS link (\emph{forward}), and GN${\rightarrow}$UAV${\rightarrow}$BS link (\emph{decode-and-forward}, with the UAV relay stationed above the BS or the GN).}}
    \label{F1andF2}
\end{figure}

Consider the deployment scenario depicted in Fig.~\ref{F1}: a swarm of $N_{U}$ rotary-wing Unmanned Aerial Vehicles (UAVs) operate as cellular relays to supplement a terrestrial Base Station (BS) by relaying data traffic dynamically generated by Ground Nodes (GNs). The BS is located at the center of the circular cell (of radius $a$), at height $H_{B}$. The UAVs operate at a fixed height $H_{U}$. The GNs are distributed uniformly at random throughout the cell, with density $\lambda_{G}$ [GNs/unit area]. \hlt{Multi-user communication is enabled via OFDMA over a spectrum of bandwidth $W$, discretized into $N_{C}$ orthogonal data channels (possibly, obtained by grouping multiple subcarriers together), each with bandwidth $B{\triangleq}\frac{W}{N_{C}}$.} We assume the system operates in the uplink, i.e., traffic requests generated by the GNs are transmitted to the BS, either directly or by using one UAV as a relay. It can be extended to both uplink/downlink via a state variable differentiating between the two.

\phantomsection\label{comm_model_label}
\noindent{\textbf{Communication Model}}: Each GN generates uplink transmission requests of $L$ bits, according to a Poisson process with rate $\lambda_{R{|}G}$ [requests/GN/unit time]. Coupled with the random deployment of GNs, uplink requests arrive in time according to a Poisson process with rate $\Lambda{\triangleq}\lambda_{G}{\cdot}\lambda_{R{|}G}\pi a^{2}$ [requests/unit time] over the circular cell. Since a new request is uniformly distributed in the cell area, the position $(r,\theta)$ of the source GN---expressed in polar coordinates with respect to the BS---has angular coordinate $\theta$ uniform in $[0,2\pi)$, and radial coordinate with probability density function given by $f_{R}(r){=}\frac{2r}{a^2}\mathbb{I}(r{\leq}a)$, where $\mathbb{I}(\cdot)$ is the indicator function. 

\hlt{A fully-connected mesh network overlaying the BS and UAVs enables command-and-control using the band-edges of the allocated spectrum as control channels. Since control packets constitute short frames relative to the large GN-generated data payloads (communicated over data channels), the control operation latencies are neglected. To request uplink transmission to the BS, a GN sends a service request with its location; the BS broadcasts this \emph{need-for-service} to the UAV swarm. Next, a consensus-driven conflict resolution process occurs among the BS and all UAVs (Sec.~\ref{S5}), based on assessed delay-energy costs for this request, culminating in a scheduling decision. If direct-BS transmission is chosen, the BS chooses an available data channel, or queues the request until one becomes available (see Sec.~\ref{S5}). The BS then instructs the GN to begin direct transmission over the data channel. Otherwise, if UAV relay $i$ is selected, the new GN request is served via a \emph{Decode-and-Forward} (D\&F) strategy on an available data channel (or queued until one becomes available), as detailed in Sec.~\ref{S5}. While executing the D\&F protocol, the UAV moves along a pre-designed energy-conscious trajectory, i.e., a sequence of way-points and velocities (see Sec.~\ref{S4}). In Sec.~\ref{S5}, we also discuss a \emph{frequency reuse} mechanism to improve spectrum utilization efficiency, and a \emph{piggybacking} mechanism allowing the scheduled UAV to serve multiple requests simultaneously. As evident from this communication model, the GN$\rightarrow$BS, GN$\rightarrow$UAV, and UAV$\rightarrow$BS links must be characterized, as detailed next.}

\noindent{\textbf{A2G Channel Model}}: For a generic link, we denote the flat-fading channel coefficient as $h{\triangleq}\sqrt{\beta}g$, where $\beta$ captures the large-scale channel variations, and $g$ with $\mathbb{E}\left[|g|^2\right]{=}1$ denotes the small-scale fading component. We model the large-scale component as $\beta{=}\beta_{\mathrm{LoS}}(d){\triangleq}\beta_{0}d^{-\alpha}$ for LoS and $\beta{=}\beta_{\mathrm{NLoS}}(d){\triangleq}\kappa\beta_{0}d^{-\tilde{\alpha}}$ for NLoS links, where $\beta_{0}$ is the pathloss at a reference distance of 1 m, $2{\leq}\alpha{\leq}\tilde{\alpha}$ are the LoS and NLoS pathloss exponents, $\kappa{\in}(0,1]$ captures the additional NLoS attenuation, and $d$ denotes the Tx-Rx Euclidean distance \cite{SCA}. Following \cite{LAP}, we use a probabilistic LoS model, with LoS probability
$
	P_{\mathrm{LoS}}(\varphi){=}[ 1{+}z_{1}\exp\{-z_{2}(\varphi{-}z_{1})\}]^{-1},
$
where $\varphi{\in}(0^{o},90^{o}]$ is the Tx-Rx elevation angle, and $z_{1}$, $z_{2}$ are parameters specific to the propagation environment (e.g., urban, suburban, rural) \cite{LAP}. The distribution of the small-scale fading component $g$ also depends on the LoS or NLoS link state \cite{WCBook}: for LoS links, as in \cite{Rician}, we model $g$ as Rician fading with a $\varphi$-dependent $K$-factor $K(\varphi){=}k_{1}\exp\{k_{2}\varphi\}$, where  $k_{1}$, $k_{2}$ are specific to the propagation environment; for NLoS links, we model $g$ as Rayleigh fading (Rician with $K{=}0$) \cite{WCBook}. Given $h$, the link capacity is $C(h){=}B{\cdot}\log_{2}\left(1{+}\frac{|h|^{2}P}{N_{0}B}\right)$, where $P$ is the transmission power, $N_{0}$ is the noise power spectral density at the receiver, and $B$ is the channel bandwidth. We assume that other sources of signal degradation, such as the Doppler effect, are well-compensated at the receiver (for example, see the approaches in ~\cite{Doppler}).

Since the large-scale fading components typically vary slowly relative to the acquisition rate of Channel State Information (CSI), we assume that the current large-scale parameters $(\beta,K)$ are known at the transmitter's side throughout the communication process, using CSI feedback over the control channel. Conversely, small-scale fading conditions vary at a much faster timescale and cannot be tracked at the transmitter. 
Hence, given $(\beta,K)$ and a transmission rate $\Upsilon$ [bits/second], we define the outage probability as $P_{\mathrm{out}}(\Upsilon,\beta,K){\triangleq}\mathbb{P}(C(\sqrt{\beta}g){<}\Upsilon)|\beta,K){=}\mathbb{P}\left(|g|^{2}{<}u(\Upsilon,\beta)\right)$, where $u(\Upsilon,\beta){\triangleq}\frac{N_{0}B}{\beta P}(2^{\frac{\Upsilon}{B}}{-}1)$. The expected throughput is then $R(\Upsilon,\beta,K){=}\Upsilon{\cdot}\left(1{-}P_{\mathrm{out}}(\Upsilon,\beta,K)\right)$, assuming that the small-scale fading is averaged out across time. The rate $\Upsilon$ is then selected to maximize the expected throughput (as opposed to the approach in \cite{Rician}, which imposes an outage probability constraint) as $\Upsilon^{*}(\beta,K){\triangleq}\argmax_{\Upsilon{\geq}0}R(\Upsilon,\beta,K)$, solved in Proposition \ref{P1}.
\hlt{\begin{prop}\label{P1}
    Given the large-scale parameters $(\beta,K)$ and $\gamma{\triangleq}\frac{N_{0}B}{\beta P}$, the optimal throughput-maximizing rate is $\Upsilon^{*}(\beta,K){=}B\log_{2}\left(1{+}\frac{Z^*}{2}\right)$, where $Z^*$ is the unique solution in $(0,\infty)$ of
    \begin{align}\label{hprime}
        h^\prime(Z) \triangleq \frac{1}{(2{+}Z)\ln\left(1{+}\frac{Z}{2}\right)} - \frac{\gamma(K{+}1)e^{-K}}{2}\frac{\exp\{-\gamma(K+1)\frac{Z}{2}\}I_{0}(\sqrt{2\gamma K(K{+}1)Z})}{Q_{1}(\sqrt{2K},\sqrt{\gamma(K{+}1)Z})} = 0,
    \end{align}
    where $I_{0}(x)$ is the modified Bessel function of first kind of order $0$, $Q_{1}(\cdot,\cdot)$ is the standard Marcum $Q$-function \cite{Rician}. $Z^*$ is solvable via the bisection method. The expected throughput is
    \begin{align}
    	R^*(\beta,K) \triangleq \max_{\Upsilon \geq 0} R(\Upsilon, \beta, K) = \Upsilon^{*}(\beta, K) \cdot Q_{1} (\sqrt{2K}, \sqrt{2(K + 1) u(\Upsilon^{*}(\beta, K),\beta)}).
    \end{align}
\end{prop}
\begin{proof}
    See Appendix~A.
\end{proof}}
\noindent{When} $K{=}0$ (Rayleigh fading for NLoS), $Q_1$ specializes to $Q_{1}(0,\sqrt{2u(\Upsilon,\beta)}){=}\exp\{-u(\Upsilon,\beta)\}$, while the condition $h^\prime(Z){=}0$ becomes $(1{+}\frac{Z}{2})\ln(1{+}\frac{Z}{2}){=}\frac{1}{\gamma}$. Finally, with the LoS and NLoS conditions averaged out in the temporal and spatial dimensions, the average link throughput is \pagebreak
\begin{align}\label{TBar}
	\bar{R}(d,\varphi) \triangleq P_{\mathrm{LoS}}(\varphi) \cdot R^{*}(\beta_{\mathrm{LoS}}(d), K(\varphi)) + (1 - P_{\mathrm{LoS}}(\varphi)) \cdot R^{*}(\beta_{\mathrm{NLoS}}(d), 0).
\end{align}
This expression is then specialized to the three distinct communication links by expressing the transmission powers, the environment-specific parameters ($z_{1}$,$z_{2}$,$k_{1}$,$k_{2}$), the large-scale parameters $(\beta,K)$, and the LoS/NLoS probabilities based on the spatial configuration, i.e., $d$ and $\varphi$. For the GN$\rightarrow$BS link, we let $\bar{R}_{GB}(r)$ be the throughput with the GN in position $(r,\theta)$, computed by setting the GN-BS distance as $d{=}\sqrt{H_{B}^{2}{+}r^{2}}$ and the elevation angle as $\varphi{=}\sin^{-1}\left(\frac{H_{B}}{d}\right)$ in \eqref{TBar}. Similarly, for the GN$\rightarrow$UAV link, we let $\bar{R}_{GU}(r_{GU})$ be the throughput when the GN-UAV distance (projected onto the $x{-}y$ plane) is $r_{GU}$, computed by setting the GN-UAV Euclidean distance as $d{=}\sqrt{r_{GU}^{2}{+}H_{U}^{2}}$ and the elevation angle as $\varphi{=}\sin^{-1}\left(\frac{H_{U}}{d}\right)$ in \eqref{TBar}. Finally, for the UAV$\rightarrow$BS link, we let $\bar{R}_{UB}(r_{UB})$ be the throughput when the $x{-}y$ projected UAV-BS distance is $r_{UB}$, computed by setting the GN-UAV Euclidean distance as $d{=}\sqrt{r_{UB}^{2}{+}(H_{U}{-}H_{B})^{2}}$ and the elevation angle as $\varphi{=}\sin^{-1}\left(\frac{H_{U}{-}H_{B}}{d}\right)$ in \eqref{TBar}. As shown in Figs.~\ref{F2}, the poor QoS experienced by GNs farther away from the BS, caused by deterioration in LoS probabilities with distance, motivates the need for UAV-relays to improve coverage throughout the cell.

\noindent{\textbf{UAV Mobility Power Model}}: For a rotary-wing UAV, since its communication power needs ($\approx$10 W) are dwarfed by its mobility power requirements ($\approx$1000 W), we model the on-board energy expenditure as a function of the horizontal flying velocity $V$ \cite{SCA}, i.e., 
\begin{align}\label{eq:Power}
    P_{\mathrm{mob}}(V) = P_{1}\left(1 + \frac{3 V^{2}}{U_{\mathrm{tip}}^{2}}\right) + P_{2}\left(\sqrt{1 + \frac{V^{4}}{4 v_{0}^{4}}} - \frac{V^{2}}{2 v_{0}^{2}}\right)^{0.5} + P_{3}V^{3},\ 0\leq V\leq V_{\max},
\end{align}
where $P_i$ are the scaling constants, $U_{\mathrm{tip}}$ is the rotor blade tip velocity, $v_{0}$ is the mean rotor induced velocity while hovering, and $V_{\max}$ is the maximum UAV flying speed \cite{SCA}.
We let $P_{\max}\triangleq \underset{0\leq V\leq V_{\max}}{\max}P_{\mathrm{mob}}(V)$ and $P_{\min}\triangleq \underset{0\leq V\leq V_{\max}}{\min}P_{\mathrm{mob}}(V)$ be the maximum and minimum power consumption of the UAV, respectively. From \cite{SCA}, hovering requires $P_{\mathrm{mob}}(0)=$1371 W, while flying at 22 m/s only consumes $P_{\min}=$936 W. This suggests that the mobility of the UAVs can be exploited to reduce power consumption, while simultaneously improving coverage across the cell. Our goal is to define an energy-conscious adaptive service scheduling and trajectory optimization scheme to minimize the time-averaged communication delay experienced by GNs in the cell, subject to an average per-UAV mobility power constraint, studied next.
\vspace{-4mm}

\section{MAESTRO: A Semi-Markov Decision Process Formulation}\label{S3}
\begin{figure} [t]
    \centering
    \includegraphics[width=0.7\linewidth]{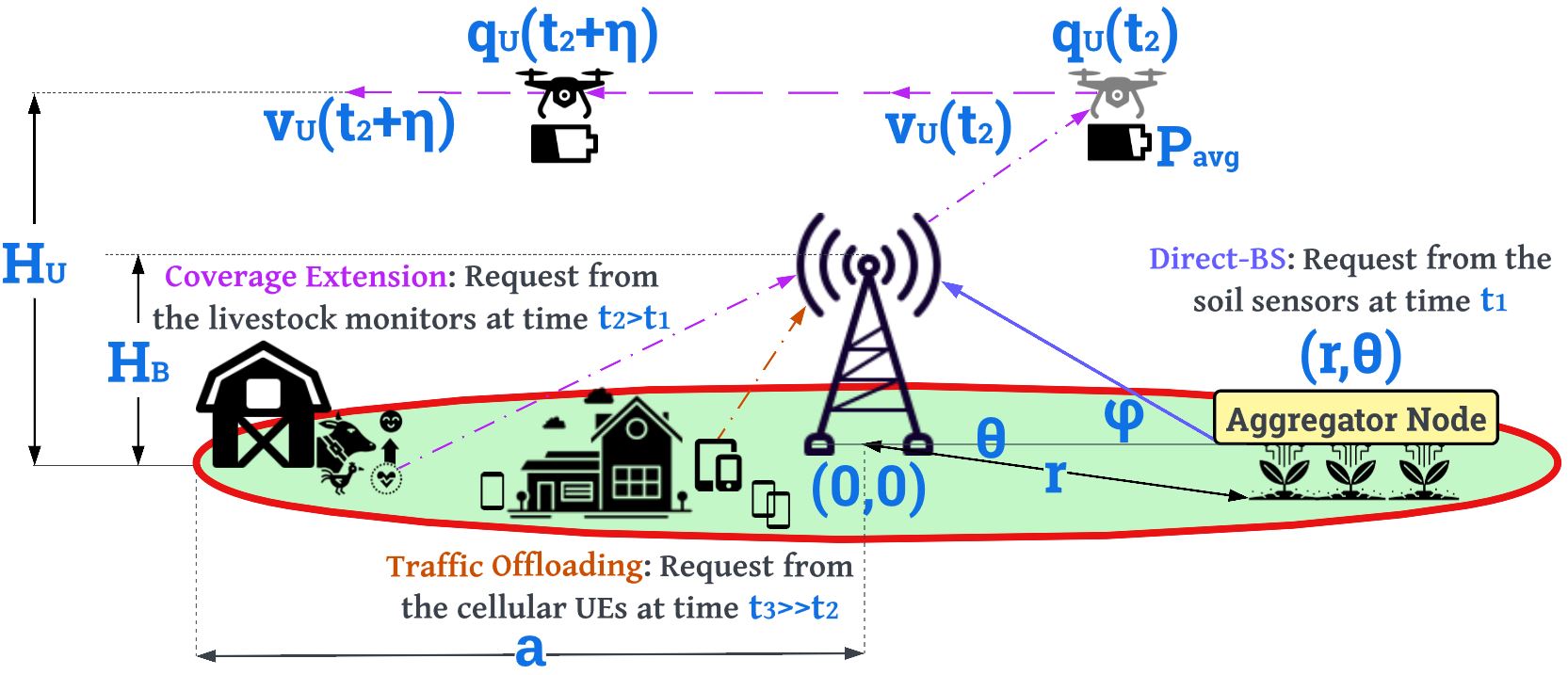}
    \vspace{-2mm}
    \caption{The single-agent specialization of our generalized deployment depicted in Fig.~\ref{F1}.}
    \label{F4}
\end{figure}

We now specialize the system model to a single UAV relay (illustrated in Fig.~\ref{F4}) via an SMDP formulation. The effective traffic rate experienced by a single UAV is $\Lambda'{\triangleq}\frac{\Lambda}{N_{U}}$ [requests/unit time/UAV], assumed in this section in place of the overall rate $\Lambda$. Let $\mathbf{q}_{U}(t){=}(r_{U}(t),\theta_{U}(t))$ be the polar coordinate of the UAV at time $t$, projected onto the $x{-}y$ plane, where $r_{U}(t){\in}\mathbb{R}_{+}$ and $\theta_{U}(t){\in}[0,2\pi)$ denote the UAV's radius and angle with respect to the BS. The system operates with the following phases. In the \emph{waiting phase}, no GN requests are being served by the UAV, which moves according to a \emph{waiting policy}. When a new GN request originates in position $(r,\theta)$, the system transitions to the \emph{request scheduling phase}, where it is determined whether the GN should transmit its data payload directly to the BS, or relay it through the UAV. In case of direct transmission, the system immediately re-enters the waiting phase, as the UAV remains free to serve other requests; else, the system enters the \emph{UAV relay phase}, in which the data payload is relayed through the UAV using the D\&F protocol; upon  completion, the system re-enters the waiting phase. \hlt{In this section, we conservatively assume that: 1) when the UAV is serving a request, it is unable to serve other incoming requests, which are thus directly served by the BS; and 2) data channels are always available at the BS to serve incoming requests. We defer to Sec.~\ref{S5} for the description of a \emph{piggybacking} mechanism to simultaneously serve multiple transmission requests, and of a queuing mechanism when data channels are unavailable.}

\noindent{\textbf{Communication Delay and UAV Energy Consumption}}: Here, we formulate the average communication delay and UAV energy consumption under a given policy $\mu$ that defines the request scheduling, communication strategy, and UAV trajectory (formally defined later). We define a decision interval as the time duration spanning the start of a waiting phase, the subsequent {request scheduling} phase when a GN request is received, until the system re-enters the waiting phase after scheduling a direct transmission to the BS, or following the {UAV relay} phase. Consider the $u$th such decision interval of duration $\Delta_{u}$, split into the time $\Delta_{u}^{(w)}$ to wait for a new request, and the time $\Delta_{u}^{(s)}$ to serve it, either through the BS (scheduling decision $\xi_{u}{=}0$) or through the UAV ($\xi_{u}{=}1$). Then, $\Delta_{u}{=}\Delta_{u}^{(w)}{+}\xi_{u}\Delta_{u}^{(s)}$, since the UAV enters the {waiting phase} immediately (and the decision interval terminates) in case of direct-BS transmission. Let $N_{u}{\geq}0$ be the number of additional requests received during the UAV relay phase of the $u$th decision period: since these are served directly by the BS, we denote their delays as $\Delta_{u,i}^{(bs)},i{=}\{1,2,{\dots},N_{u}\}$. Let $E_{u}$ be the UAV mobility energy expended during the $u$th decision interval, and let $M_{t}$ be the total number of decision intervals completed up to time $t$. We define the expected long-term average communication delay per request ($\bar{D}_{\mu}$) and average UAV power ($\bar{P}_{\mu}$), under $\mu$, as
\begin{align}\label{eq:DBarMu}
    &\bar{D}_{\mu} \triangleq \lim_{t \rightarrow \infty} \mathbb{E}_{\mu} \Bigg[\frac{\frac{1}{M_t}\sum_{u=1}^{M_t}(\Delta_u^{(s)} + \xi_u\sum_{i=1}^{N_u}\Delta_{u, i}^{(bs)})}{\frac{1}{M_t}\sum_{u=1}^{M_t}(1 + \xi_u N_u)}\Bigg],\ \bar{P}_{\mu} \triangleq \lim_{t \rightarrow \infty} \mathbb{E}_{\mu} \Bigg[ \frac{\frac{1}{M_t}\sum_{u=1}^{M_t}E_u}{\frac{1}{M_t}\sum_{u=1}^{M_t}\Delta_u}\Bigg].
\end{align}
Note that $\bar{D}_{\mu}$ in \eqref{eq:DBarMu} captures the delays of all requests, i.e., those relayed through the UAV ($\xi_{u}{=}1$), those transmitted directly to the BS ($\xi_{u}{=}0$), as well as the $N_{u}$ additional requests served directly by the BS during the UAV relay phase. \hlt{Thus, the objective is to solve 
\begin{align}\label{eq:opt_prob_orig}
    \bar D^* = \min_{\mu} \, \bar{D}_{\mu}, \; \mathrm{s.t.} \; \bar P_{\mu} \leq P_{\mathrm{avg}},
\end{align}
where $P_{\mathrm{avg}}\in(P_{\min},P_{\max})$ is the average power constraint,
and the optimal policy is denoted as $\mu^*$.} To simplify, let $\bar{\mathbb{E}}_{\mu}[C_{u}]{\triangleq}\lim\limits_{t{\rightarrow}\infty}\mathbb{E}_{\mu}[\frac{1}{M_{t}}\sum_{u{=}1}^{M_{t}}C_{u}]$ be a shorthand notation for the long-term average cost $C_u$ per decision interval. Let $\bar{E}_{\mu}{\triangleq}\bar{\mathbb{E}}_{\mu}\left[E_{u}\right]$ be the average UAV energy expenditure, $\bar{T}_{\mu}{\triangleq}\bar{\mathbb{E}}_{\mu}\left[\Delta_{u}\right]$ be the average interval duration, $\bar{N}_{\mu}{\triangleq}\bar{\mathbb{E}}_{\mu}[1+\xi_{u}N_{u}]$ be the average number of requests served, $\bar{W}_{\mu}^{(s)}{\triangleq}\bar{\mathbb{E}}_{\mu}[\Delta_{u}^{(s)}]$ be the average delay of requests for which a scheduling decision is made, $\bar{W}_{\mu}^{(bs)}{\triangleq}\bar{\mathbb{E}}_{\mu}[\xi_{u}\sum_{i{=}1}^{N_{u}}\Delta_{u,i}^{(bs)}]$ be the average delay of requests served directly by the BS during the UAV relay phase, per decision interval. \hlt{Using Little's Law \cite{LittlesLaw}, we  can then express $\bar{P}_{\mu}{=}\frac{\bar{E}_{\mu}}{\bar{T}_{\mu}}$ and $\bar{D}_{\mu}{=}\frac{\bar{W}_{\mu}^{(s)}+\bar{W}_{\mu}^{(bs)}}{\bar{N}_{\mu}}$, hence the optimization problem can be recast as
\begin{align}\label{eq:OverallObj1}
    \bar{D}^{*} = \underset{\mu}{\mathrm{min}} \; \frac{\bar{W}_{\mu}^{(s)} + \bar{W}_{\mu}^{(bs)}}{\bar{N}_{\mu}} \; \mathrm{s.t.} \; 
    \bar{\mathcal E}_\mu\triangleq
    \bar{E}_{\mu} - P_{\mathrm{avg}}\bar{T}_{\mu} \leq 0,
\end{align}
where $\bar{\mathcal E}_\mu{=}\bar{\mathbb{E}}_{\mu}[E_{u}{-}P_{\mathrm{avg}}\Delta_u]$ is the \emph{excess energy cost}.
Note the inherent complexity to solve \eqref{eq:OverallObj1}: as the policy varies, the delay metric changes both the numerator and denominator of the objective function, precluding a direct application of dynamic programming tools.}

\noindent{\textbf{\hlt{Alternative Problem Formulation}}}:\label{altopt} \hlt{
To address this challenge, we now devise a surrogate optimization metric, by characterizing upper and lower bounds to $\bar D_{\mu}$. To this end, let us define a "baseline" policy $\mu_{BS}$ as the one such that all requests are served by the BS and the UAV flies around at minimum power $P_{\min}$ (this policy is feasible). Since the delay to serve a request from a GN in position $(r,\theta)$ by direct transmission to the BS is $\frac{L}{\bar R_{GB}(r)}$,  the expected delay under policy $\mu_{BS}$ is obtained by computing the expectation with respect to the radial coordinate, $\bar{D}_{BS}{\triangleq}\int_{0}^{a}\frac{L}{\bar{R}_{GB}(r)}f_{R}(r)\mathrm{d}r$. Clearly, optimization of the policy yields $\bar{D}^{*}{\leq}\bar{D}_{BS}$. Under any policy $\mu$ (including $\mu^*$) better than 
$\mu_{BS}$ (i.e., such that $\bar{D}_{\mu}\leq\bar{D}_{BS}$), the following bounds hold.
\begin{prop}\label{P2}
    Let $\mu$ be such that $\bar{D}_{\mu}\leq\bar{D}_{BS}$. Then, it holds that
    \begin{align}\label{bounds}
        \bar{W}_{\mu}^{(s)} \leq \bar{D}_{\mu} \leq \bar{W}_{\mu}^{(s)} \frac{1 + \Lambda' \bar{D}_{BS}}{1 + \Lambda' \bar{W}_{\mu}^{(s)}} \leq \bar{D}_{BS}.
    \end{align}
\end{prop}
\begin{proof}
    See Appendix~B.
\end{proof}
\noindent{Noticing} that both the lower and upper bounds of $\bar{D}_{\mu}$ are increasing functions of $\bar{W}_{\mu}^{(s)}$, in our subsequent analyses we will focus on the alternative  optimization problem
\begin{align}
\label{minW}
   \underset{\mu}{\mathrm{min}} \; \bar{W}_{\mu}^{(s)} \; \mathrm{ s.t. } \; \bar{\mathcal E}_{\mu} \leq 0.
\end{align}
In Sec. \ref{S6} (see Table \ref{T4}), we show that this alternative formulation leads to a near-optimal solution with respect to the original optimization \eqref{eq:opt_prob_orig}.}
To solve \eqref{minW}, we define the Lagrangian
\begin{align}\label{eq:W2Lagr}
    g(\nu) = \underset{\mu}{\mathrm{min}}\, \bar{W}_{\mu}^{(s)} + \nu\bar{\mathcal E}_{\mu} = \underset{\mu}{\mathrm{min}}\, \lim_{t\rightarrow\infty} \, \mathbb{E}_{\mu} \left[\frac{1}{M_t} \sum_{u=1}^{M_t}\left(\Delta_u^{(s)} + \nu (E_u{-}P_{\mathrm{avg}}\Delta_u)\right)\right],
\end{align}
where $\nu$ is the dual variable, optimized by solving $\max_{\nu{\geq}0}g(\nu)$. We now demonstrate that for a given $\nu{\geq}0$, \eqref{eq:W2Lagr} can be cast as a Semi-Markov Decision Process (SMDP) and solved with dynamic programming tools. Next, we discuss the SMDP states, actions, transitions, and policy. 

\noindent{\textbf{States}}: The state is defined by the UAV position $\mathbf{q}_{U}$, an element of the set $\mathcal{Q}_{\mathrm{UAV}}{\triangleq}\mathbb{R}_{+}{\times}[0,2\pi)$ (polar coordinates), and the position $\mathbf{q}_{G}$ of the GN originating traffic, taking values from the set  $\mathcal{Q}_{\mathrm{GN}}{\triangleq}[0,a]{\times}[0,2\pi)$. The state space is then $\mathcal{S}{=}\mathcal{S}_{\mathrm{wait}}\cup\mathcal{S}_{\mathrm{comm}}$, where $\mathcal{S}_{\mathrm{wait}}{=}\mathcal{Q}_{\mathrm{UAV}}$ is the set of \emph{waiting} states and $\mathcal{S}_{\mathrm{comm}}{=}\mathcal{Q}_{\mathrm{UAV}}{\times}\mathcal{Q}_{\mathrm{GN}}$ is the set of \emph{communication} states. Crucial to the definition of the SMDP is how the system is sampled in time to define Markovian dynamics in the evolution of the sampled states: accordingly, we define the actions available in each state $\mathbf{s}{\in}\mathcal{S}$ and the transition probabilities, along with the time duration $T(\mathbf{s};\mathbf{a})$, the UAV energy usage $E(\mathbf{s};\mathbf{a})$, and the request service delay $\Delta(\mathbf{s};\mathbf{a})$ metrics accrued in state $\mathbf{s}$ under action $\mathbf{a}$.

\noindent{\textbf{Waiting states' actions, transitions, and metrics}}: In waiting state $\mathbf{s}{=}\mathbf{q}_{U}{\in}\mathcal{S}_{\mathrm{wait}}$ at time $t$, i.e., the UAV is in position $\mathbf{q}_{U}(t){=}\mathbf{q}_{U}{=}(r_{U},\theta_{U})$ with no active requests, then the UAV moves with radial and angular velocity components $(v_{r},\theta_{c})$, over an arbitrarily small duration $\Delta_{0}{\ll}\frac{1}{\Lambda'}$. Thus, the waiting-state action space is $\mathcal{A}_{\mathrm{wait}}(r_{U}){\triangleq}\Big\{(v_{r},\theta_{c}){\in}\mathbb{R}^{2}\Big|\sqrt{v_{r}^{2}{+}r_{U}^{2}{\cdot}\theta_{c}^{2}}{\leq}V_{\mathrm{max}} \Big\}$, where $v_{U}{=}\sqrt{v_{r}^{2}{+}r_{U}^{2}\theta_{c}^{2}}$ is the velocity expressed using polar coordinates. Upon choosing action $\mathbf{a}{=}(v_{r},\theta_{c}){\in}\mathcal{A}_{\mathrm{wait}}(r_{U})$, the communication delay is $\Delta(\mathbf{s};\mathbf{a}){=}0$, since there is no ongoing communication; the duration of a waiting state is $T(\mathbf{s};\mathbf{a}){=}\Delta_{0}$, and the UAV's energy use is $E(\mathbf{s};\mathbf{a}){=}\Delta_{0}P_{\mathrm{mob}} \left(v_{U}\right)$ to move at velocity $v_{U}$. The new state is then sampled at time $t{+}\Delta_{0}$, with the UAV moved to the new position $\mathbf{q}_{U}(t{+}\Delta_{0}){\approx}(r_{U},\theta_{U}){+}(v_{r},\theta_{c})\Delta_{0}$. With probability $e^{-\Lambda'\Delta_{0}}$, no new request is received in the time interval $[t,t{+}\Delta_{0}]$, so that the new state is a waiting state.  Otherwise, a new request is received from a GN in position $(r,\theta)$ (communication state). The transition probabilities from the waiting state $\mathbf{s}_{n}{=}\mathbf{q}_{U}{\in}\mathcal{S}_\mathrm{wait}$ under action $\mathbf{a}_{n}{=}(v_{r},\theta_{c}){\in}\mathcal{A}_{\mathrm{wait}}(r_{U})$ are thus
\begin{align}\label{eq:CommTransProb}
    &\mathbb{P}(\mathbf{s}_{n+1} = \mathbf q_U + \mathbf a_n\Delta_0 | \mathbf{s}_n,\mathbf{a}_n) = e^{-\Lambda'\Delta_{0}},\\\nonumber
    &\mathbb{P}(\mathbf{s}_{n+1} = (\mathbf q_U + \mathbf a_n\Delta_0,\mathbf q_G') \text{ with } \mathbf q_G' \in \mathcal{F} \,|\mathbf{s}_n,\mathbf{a}_n) = \frac{A(\mathcal{F})}{\pi a^2} \cdot (1 - e^{-\Lambda'\Delta_{0}}),\ \forall \mathcal{F}\subseteq \mathcal{Q}_{\mathrm{GN}},
\end{align}
where $A(\mathcal{F})$ is the area of region $\mathcal{F}$, since requests are uniformly distributed in the cell.

\noindent{\textbf{Communication states' actions, transitions, and metrics}}: Upon reaching a communication state $\mathbf{s}_{n}{=}(\mathbf{q}_{U},\mathbf{q}_{G}){\in}\mathcal{S}_{\mathrm{comm}}$ at time $t$, the system must serve a GN request at position $\mathbf{q}_{G}{=}(r,\theta)$. The BS first determines the scheduling decision $\xi{\in}\{0,1\}$. If $\xi{=}0$,
 denoted as the action $\mathbf a{=}\mathrm{BS}$,
 the GN transmits directly to the BS; the next state is 
the waiting state $\mathbf{s}_{n{+}1}{=}\mathbf{q}_{U}$, sampled immediately after, resulting in the energy-time metrics
$E(\mathbf{s}_{n};\mathbf{a}){=}T(\mathbf{s}_{n};\mathbf{a}){=}0$, and service delay metric $\Delta(\mathbf{s}_{n};\mathbf{a}){=}\frac{L}{\bar R_{GB}(r)}$ (time required to transmit the payload with throughput $\bar R_{GB}(r)$ between the GN and the BS).
Instead, if $\xi{=}1$, the UAV uses the D\&F protocol, while following a trajectory starting from its current position $\mathbf{q}_{U}$ and ending in position $\mathbf{q}_{U}'$. We denote this action as $\mathbf a{=}(\mathbf q_U{\rightarrow}\mathbf q_{U}')$. In the \emph{decode} phase of D\&F (of duration $t_{p}$), the GN transmits its data payload to the UAV; in the \emph{forward} phase (of duration $\Delta{-}t_{p}$), the UAV relays it to the BS. Assuming a \emph{move-and-transmit} strategy \cite{SCA}, the trajectory ($\mathbf{q}_{U}{\rightarrow}{\mathbf{q}}_{U}'$) and the durations ($t_{p}$ and $\Delta{-}t_{p}$) must satisfy the data payload constraints \eqref{eq:PLConst1}, i.e., the entire payload of $L$ bits is first transmitted to the UAV with throughput $\bar R_{GU}(r_{GU}(\eta))$, and then relayed to the BS
with throughput $\bar{R}_{UB}(r_{UB}(\eta))$, where
 $r_{GU}({\eta})$  and $r_{UB}(\eta)$ are the GN-UAV and UAV-BS distances (projected onto the $x{-}y$ plane) at time $\eta$ along the trajectory, respectively, so that the total communication delay is $\Delta$. For this action, the cost metrics are $\Delta(\mathbf{s}_{n};\mathbf{{a}}){=}T(\mathbf{s}_{n};\mathbf{{a}}){=}\Delta$ and $E(\mathbf{s}_{n};\mathbf{{a}}){=}\int_{0}^{\Delta}P_{\mathrm{mob}}\left(v_{U}(\eta)\right)\mathrm{d}\eta$. Upon completing D\&F at time $t{+}\Delta$, the UAV enters the waiting state ($\mathbf{s}_{n{+}1}{=}\mathbf{q}_{U}'$). The set of feasible UAV trajectories from $\mathbf{q}_{U}$ to $\mathbf{q}_{U}'$, to serve a GN at position $\mathbf{q}_{G}$ is
\begin{align}
\label{eq12}
&	\mathcal{Q}_{\mathbf q_G} \big({\mathbf q}_U\rightarrow{\mathbf q}_U'\big) \triangleq \Big\{ \mathbf{p}_{U} : [0,\Delta] \mapsto \mathbb{R}_{+} \times[0,2\pi)\text{ s.t.}\\
	& \;\;\;\; \int_{0}^{t_{p}} \bar{R}_{GU}(r_{GU}(\eta)) \mathrm d \eta \geq L, \ \int_{t_p}^{\Delta} \bar R_{UB}(r_{UB}(\eta)) \mathrm d \eta \geq L, \label{eq:PLConst1}\tag{C.1}\\
	& \;\;\;\; v_U (\eta) \leq V_{\mathrm{max}},\ 
\mathbf{p}_{U}(0) ={\mathbf q}_U, \mathbf{p}_{U}(\Delta) ={\mathbf q}_U',\label{eq:IFConst1}\ \exists \Delta \geq 0, \exists\; 0 \leq t_p \leq \Delta \Big\},\tag{C.2}
\end{align}
 where $v_U(\eta)$ is the UAV speed, \ref{eq:PLConst1} reflects the data payload constraints, and \ref{eq:IFConst1} the maximum speed and trajectory constraints. Then, the action space in state $(\mathbf{q}_{U},\mathbf{q}_{G}){\in}\mathcal{S}_{\mathrm{comm}}$ when $\xi{=}1$ is the set $\mathcal{Q}_{\mathbf{q}_{G}}(\mathbf{q}_{U}){\triangleq}\cup_{\mathbf{q}_{U}'{\in}\mathcal{Q}_{\mathrm{UAV}}}\mathcal{Q}_{\mathbf{q}_{G}}\big(\mathbf{q}_{U}{\rightarrow}\mathbf{q}_{U}'\big)$ of feasible trajectories starting in $\mathbf{q}_{U}$ that serve the GN at $\mathbf{q}_{G}$ via the D\&F protocol. The overall action space of this communication state is then $\mathcal{A}_{\mathrm{comm}}(\mathbf{q}_{U},\mathbf{q}_{G}){\triangleq}\{\mathrm{BS}\}{\cup}\{\mathcal{Q}_{\mathbf{q}_{G}}(\mathbf{q}_{U})\}$, including the scheduling decision $\xi\in\{0,1\}$.

\noindent{\textbf{Policy $\mu$}}: For waiting states $\mathbf{q}_{U}{\in}\mathcal{S}_{\mathrm{wait}}$, the policy  $\mu(\mathbf{q}_{U}){\in}\mathcal{A}_{\mathrm{wait}}(r_{U})$ selects a velocity $(v_{r},\theta_{c})$ from the respective action space. Likewise, for communication states $(\mathbf{q}_{U},\mathbf{q}_{G}){\in}\mathcal{S}_{\mathrm{comm}}$, the policy selects the scheduling decision $\xi{\in}\{0,1\}$ and if $\xi{=}1$, the trajectory followed in the D\&F protocol, i.e., $\mu(\mathbf{q}_{U},\mathbf{q}_{G}){\in} \mathcal Q_{\mathbf q_G} (\mathbf q_U)$. With a stationary policy $\mu$ defined, the Lagrangian metric $L_{\mu}^{(\nu)}{\triangleq}\bar{W}_{\mu}^{(s)}{+}\nu\bar{\mathcal E}_{\mu}$ in \eqref{eq:W2Lagr} is reformulated using Little's Law \cite{LittlesLaw} and is written as 
\begin{align}\label{eq:CostMetric}
    L_\mu^{(\nu)} = \lim_{N \rightarrow \infty} \mathbb{E}_\mu \Bigg[ \frac{\frac{1}{N}\sum_{n=0}^{N-1}  \ell_\nu(\mathbf{s}_n; \mu(\mathbf{s}_n)) }{\frac{1}{N}\sum_{n = 0}^{N-1} \mathbb I(\mathbf{s}_n \in \mathcal{S}_{\mathrm{comm}})}  \Bigg] = \frac{1}{\pi_{\mathrm{comm}}}\int_{\mathcal{S}} \Pi_{\mu}(\mathbf{s})\ell_\nu(\mathbf{s}; \mu(\mathbf{s}))\mathrm{d}\mathbf{s},
\end{align}
where $\Pi_{\mu}(\mathbf{s})$ is the steady-state probability density function of being in state $\mathbf{s}$ under policy $\mu$, $\pi_{\mathrm{comm}}{=}\int_{\mathcal{S}_{\mathrm{comm}}}\!\!\!\!\!\Pi_{\mu}(\mathbf{s})\mathrm{d}\mathbf{s}$ is the steady-state probability that the UAV is in the communication phase, and $\ell_{\nu}(\mathbf{s};\mathbf{a}){\triangleq}\Delta(\mathbf{s};\mathbf{a}){+}\nu\big(E(\mathbf{s};\mathbf{a}){-}P_{\mathrm{avg}}T(\mathbf{s};\mathbf{a})\big)$ is the Lagrangian metric in state $\mathbf{s}$ under action $\mathbf{a}$. In \eqref{eq:CostMetric}, $\sum_{n=0}^{N{-}1}\ell_{\nu}(\mathbf{s}_{n};\mu(\mathbf{s}_{n}))$ is the total Lagrangian cost accrued during the first $N$ SMDP stages, and $\sum_{n{=}0}^{N{-}1}\mathbb{I}(\mathbf{s}_{n}{\in}\mathcal{S}_{\mathrm{comm}})$ is the number of communication states encountered; since a new decision interval initiates after a communication state, this equals the number of decision intervals ($M_t$ in \eqref{eq:W2Lagr}). Taking the limit $N{\to}\infty$, $L_{\mu}^{(\nu)}$ is the expected Lagrangian cost per decision interval, as expressed in \eqref{eq:W2Lagr}. The right-hand side expression in \eqref{eq:CostMetric} follows because the SMDP reaches the steady-state when $N{\to}\infty$. Specializing, $\ell_{\nu}(r_{U},\theta_{U};v_{r},\theta_{c}){=}\nu(P_{\mathrm{mob}}(\sqrt{v_{r}^{2}{+}r_{U}^{2}\theta_{c}^{2}}){-}P_{\mathrm{avg}})\Delta_{0}$ for the waiting states, $\ell_{\nu}(r_{U},\theta_{U},r,\theta; \mathrm{BS}){=}\frac{L}{\bar{R}_{GB}(r)}$ for direct-BS transmission in communication states, and $\ell_{\nu}(r_{U},\theta_{U},r,\theta;\mathbf{p}_{U}){=}(1{-}\nu P_{\mathrm{avg}})\Delta{+}\nu\int_{0}^\Delta P_{\mathrm{mob}}\left(V(\eta)\right)\mathrm{d}\eta$ for a communication  relayed through the UAV.
The next proposition shows that the steady-state probability $\pi_{\mathrm{comm}}$ is independent of the policy $\mu$, i.e., it is not affected by the optimization over $\mu$.
\hlt{\begin{prop}\label{P3}
We have
$\pi_{\mathrm{comm}}{=}1-(2{-}e^{-\Lambda'\Delta_{0}})^{-1}$.
\end{prop}
\begin{proof}    See Appendix~C.\end{proof}
}
\noindent{This} result permits rewriting \eqref{eq:W2Lagr} as an \emph{average cost-per-stage problem}
\begin{align}\label{eq:TotalGMin}
	g(\nu) = \frac{1}{\pi_{\mathrm{comm}}}\underset{\mu}{\mathrm{min}} \; \int_{\mathcal{S}} \Pi_{\mu}(s) \ell_\nu(s; \mu(s))\mathrm d s,
\end{align}
solvable through standard dynamic programming approaches (upon discretization of the state and action spaces), followed by the dual maximization $\mathrm{max}_{\nu{\geq}0}g(\nu)$.

\noindent{\textbf{Two-stage policy decomposition}}: Since GN transmission requests are uniformly distributed in the circular cell, the UAV radius is a sufficient statistic in decision-making for a waiting state $(r_{U},\theta_{U})$, expressed as $r_{U}{\in}\mathcal{S}_{\mathrm{wait}}\triangleq[0,a]$. Likewise, for a communication state $(r_{U},\theta_{U},r,\theta)$, only the UAV radius, GN request radius, and the angle $\psi{\in}[0,2\pi)$ between them suffice to characterize the state. Thus, communication states can be compactly represented as $(r_{U},r,\psi{=}\theta{-}\theta_U){\in}\mathcal{S}_{\mathrm{comm}}\triangleq[0,a]^2{\times}[0, 2\pi)$. Hence, the policy affects the SMDP state transitions (and its steady-state) only through the UAV radial velocity $v_{r}$ in the waiting states, the
scheduling decision (direct-BS or UAV relay) and
 UAV trajectory's end radius position $\hat{r}_{U}$ in communication states. Instead, the angular velocity $\theta_{c}$ in the waiting states and the UAV trajectory to reach the target end radius $\hat{r}_{U}$ in the communication states
  only affect the instantaneous Lagrangian $\ell_{\nu}$, but not
 state dynamics.

With this observation, let $O(r_U){\triangleq}v_{r}{\in}[-V_{\mathrm{max}},V_{\mathrm{max}}]$ define the radial velocity policy of waiting states $r_U{\in}\mathcal{S}_{\mathrm{wait}}$, specifying the radial velocity component of waiting action $(v_r,\theta_c){\in}\mathcal{A}_{\mathrm{wait}}(r_U)$; 
let $U(r_U,r,\psi){\triangleq}(\xi,\hat{r}_{U})$ define the scheduling and next radius position policy of communication states $(r_U,r,\psi){\in}$ $\mathcal{S}_{\mathrm{comm}}$:
either direct-BS with $\hat r_U=r_U$ ($\xi=0$), or any trajectory
starting from radius $r_U$ and ending at radius $\hat r_U$ when relaying through the UAV ($\xi=1$). Accordingly, $O$ and $U$ are the SMDP's \emph{outer decisions} and are the only actions affecting the steady-state distribution, denoted as $\Pi_{O,U}$ under the outer policy $(O,U)$; thus, \eqref{eq:TotalGMin} can be restated as
\begin{align}\label{eq:PolDecomp}
	g(\nu) = \frac{1}{\pi_{\mathrm{comm}}} \underset{O,U}{\mathrm{min}} \Bigr[ \int_{\mathcal{S}_{\mathrm{wait}}} \Pi_{O,U}(\mathbf{s}) \ell_{\nu}^{*}(\mathbf{s}; O(\mathbf{s}))\mathrm{d}\mathbf{s} + \int_{\mathcal{S}_{\mathrm{comm}}} \Pi_{O,U}(\mathbf{s}) \ell_{\nu}^{*}(\mathbf{s}; U(\mathbf{s})) \mathrm{d}\mathbf{s} \Bigr],
\end{align}
where $\ell_{\nu}^{*}$ is the Lagrangian metric optimized with respect to the \emph{inner decision} components not specified by $O$ and $U$. In particular, for a waiting state $r_{U}$, under the radial velocity action $O(r_U){=}v_{r}$, the inner optimization is performed with respect to the angular velocity $\theta_{c}$,\pagebreak
\begin{align}\label{eq:MinLWP}
	&\ell_{\nu}^{*}(r_U; v_r) = \underset{\theta_c}{\mathrm{min}}\; \nu \left( P_{\mathrm{mob}}(V) - P_{\mathrm{avg}} \right)\Delta_0 \;\; \mathrm{s.t.}\;\; V=\sqrt{v_{r}^{2} + r_U^2\theta_c^2} \leq V_{\mathrm{max}}.
\end{align}
Since $\nu{\geq}0$, the optimizer $\theta_{c}^{*}$ is the angular velocity minimizing the UAV power consumption: due to the quasi-convex structure of $P_{\mathrm{mob}}(v)$ \cite{SCA}, $\theta_{c}^{*}{=}0$ if $|v_r|{\geq}v_{P_{\mathrm{min}}}{\triangleq}\arg\min_V P_{\mathrm{mob}}(V)$ (in fact, any angular movement would undesirably increase power consumption), and $\sqrt{v_{r}^{2}{+}r_{U}^{2}(\theta_{c}^{*})^{2}}{=}v_{P_{\mathrm{min}}}$ otherwise (i.e., enough angular movement to yield the power minimizing speed). For communication states, under direct-BS transmission,
 $\ell_{\nu}^{*}(\mathbf{s};0,r_U)=L/R_{GB}(r)$; on the other hand, when relaying through the UAV, $\ell_{\nu}^{*}$
is obtained by optimizing the trajectory $\mathbf{p}_{U}$ followed by the UAV, starting at radius $r_U$ and terminating at radius $\hat{r}_{U}$ (with final angular position $\hat\phi$ optimized),
\begin{align}
    \ell_{\nu}^{*}(\mathbf{s};1,\hat{r}_{U}){=}&\underset{\Delta,\mathbf{p}_{U},t_{p},\hat\phi}{\mathrm{min}}(1{-}\nu P_{\mathrm{avg}})\Delta{+}\nu\int_{0}^{\Delta}P_{\mathrm{mob}}(v_U(\eta))\mathrm{d}\eta \text{ s.t.}\;\, \text{\ref{eq:PLConst1}},\text{\ref{eq:IFConst1}}\label{eq:EllMinold}.
\end{align}
where \ref{eq:PLConst1}-\ref{eq:IFConst1} are the data payload, maximum UAV speed and trajectory constraints (see \eqref{eq12}).
 In other words, the inner decision on trajectory minimizes the instantaneous delay-energy trade-off, among all feasible trajectories terminating at the target radius $\hat{r}_{U}$. 
\hlt{Defining $\alpha{\triangleq}\frac{\nu P_{\max}}{(1{+}\nu(2P_{\max}{-}P_{\mathrm{avg}}))} \in [0,1]$ to regulate the trade-off between service delay and UAV energy, \eqref{eq:EllMinold} can be rewritten as
\begin{align}
    \frac{\ell_{\nu}^{*}(\mathbf{s};1,\hat{r}_{U})}{1{+}\nu(2P_{\max}{-} P_{\mathrm{avg}})}{=}&
    \underset{\Delta,\mathbf{p}_{U},t_{p}}{\mathrm{min}}(1-2\alpha)\Delta
    {+}\alpha\int_{0}^{\Delta}
\frac{P_{\mathrm{mob}}(V(\eta))
}{P_{\max}}\mathrm{d}\eta \text{ s.t.}\;\, \text{\ref{eq:PLConst1}},\text{\ref{eq:IFConst1}}\label{eq:EllMin},
\end{align}
This reformulation is the focus of our HCSO trajectory design algorithm, detailed in Sec.~\ref{S4}.}

Alg.~\ref{A1} optimizes the outer policy and computes the average cost-per-stage metric $g(\nu)$, along with the average excess energy-per-stage metric for a given $\nu$, by solving problem \eqref{eq:PolDecomp} via value iteration \cite{Bertsekas}. Alg.~\ref{A2} solves the dual maximization $\mathrm{max}_{\nu{\geq}0}g(\nu)$ via projected sub-gradient ascent\footnote{The source code for these algorithms is available on \href{https://github.com/bharathkeshavamurthy/MAESTRO-X.git}{GitHub} \cite{MAESTRO-X}.} \cite{SubgradientMethods}. Specifically, in Alg.~\ref{A1}, lines 2 and 3 compute the inner Lagrangian cost metric optimized with respect to the inner actions---along with the excess energy cost metric---for all states and outer actions; line 6 computes the value iteration update for waiting states:
upon moving to the new radial position $r_{U}{+}v_{r}\Delta_{0}$, 
no request is received, w.p. $e^{-\Lambda'\Delta_{0}}$, hence moving to a waiting state (with future value $V_{W,i}(r_{U}{+}v_{r}\Delta_{0})$); otherwise, the system moves to a communication state, with future value $V_{C,i}(r_{U}{+}v_{r}\Delta_{0})$ (averaged with respect to the request position); line 12 computes the value iteration update for communication states, transitioning to a waiting state w.p. 1; the corresponding optimal outer actions are saved in lines 7 and 13; line 16 averages the value of communication states with respect to the random request position; lines 8, 14, and 17 similarly update the total excess energy cost, needed to compute the projected dual sub-gradient ascent in Alg.~\ref{A2}.
\label{discretizeVI}\hlt{In practice, the integrals in lines 16 and 17, and the continuous state/action spaces are discretized (see MAESTRO-X \cite{MAESTRO-X}), leading to an overall complexity of each value iteration update (lines 5-18) of order $\mathcal O(K_R\cdot(K_V+K_R^2\cdot K_A))$, where $K_R$ is the number of discretized radii levels ($r_U$ and $r$ values), $K_A$ is the number of angular levels ($\psi$ and $\psi'$), and $K_V$ is the number of discretized radial velocities ($v_r$).} Upon convergence \hlt{(typically, value iteration converges within $\mathcal O(\log(1/\delta))$ iterations to achieve a target accuracy $\delta$ \cite[Sec. V]{Bertsekas})}, line 21 estimates the values of the average cost-per-stage and excess energy-per-stage metrics.

In Alg.~\ref{A2}, line 1 initializes the dual variable and a sequence of step-sizes used for projected sub-gradient ascent; line 3 calls value iteration (Alg.~\ref{A1}) using the current dual variable $\nu$, and outputs the optimal outer policy and the average cost-, excess energy- per-stage metrics; line 5 monitors convergence in terms of  primal feasibility and complementary slackness conditions; line 4 updates the value of the dual variable in the direction of its sub-gradient and projects its value to the non-negative range to ensure dual feasibility;
note that Alg.~\ref{A1}  outputs also the \emph{relative values} metrics $V$ and $\mathcal E$:
these are used to initialize the total cost and excess energy metrics in the next call to Alg.~\ref{A1}, and help speed up convergence. We are left with the trajectory design (line $3$ of Alg.~\ref{A1}), carried out using Hierarchical CSO in the next section.
\begin{algorithm} [t]
\caption{$(O^{*},U^{*},g(\nu),\bar{\mathcal E},V_{\cdot,0}^{next},\mathcal E_{\cdot,0}^{next})=\mathrm{VITER}(\nu,V_{\cdot,0},\mathcal E_{\cdot,0})$}\label{A1}
    \begin{algorithmic}[1]
        \scriptsize
        \State \textbf{Initialization}: $i{=}0$; stop criterion $\delta$.
        \State \textbf{Inner optimization in waiting states}: ${\forall}r_U{\in}\mathcal{S}_{\mathrm{wait}}, {\forall}v_{r}{\in}[-V_{\mathrm{max}},V_{\mathrm{max}}]$, calculate $\ell_{\nu}^{*}(r_U;v_{r})$ as in \eqref{eq:MinLWP}, with minimizer $\theta_{c}^{*}$; compute excess energy cost $\epsilon^{*}(r_U;v_{r}){=}
        P_{\mathrm{mob}}(\sqrt{v_{r}^{2} + r_U^2(\theta_c^*)^2})\Delta_0-P_{\mathrm{avg}}\Delta_0$.
    	\State \textbf{Inner optimization in communication states}: ${\forall}\mathbf{s}{\in}\mathcal{S}_{\mathrm{comm}}, {\forall}\hat{r}_{U}{\in}[0,a]$, calculate $\ell_{\nu}^{*}(\mathbf{s};1,\hat{r}_{U})$ via Alg.~\ref{A3} with
	 $\alpha=\nu P_{\max}/(1{+}\nu(2P_{\max}{-} P_{\mathrm{avg}}))$, with minimizer $\mathbf{p}_{U}^{*}$ (trajectory); compute excess energy cost $\epsilon^{*}(\mathbf{s};\hat{r}_{U}){=}E(\mathbf{s};\mathbf{p}_{U}^{*})-P_{\mathrm{avg}}T(\mathbf{s};\mathbf{p}_{U}^{*})$.
    	\Repeat
            \For{each $r_{U}{\in}[0,a]$} \Comment{Outer optimization in waiting states}
        	    \State $V_{W,i{+}1}(r_{U}){\gets}\underset{v_{r}{\in}[-V_{\mathrm{max}},V_{\mathrm{max}}]}{\mathrm{min}}\big[\ell_{\nu}^{*}(r_{U};v_{r}){+}e^{-\Lambda'\Delta_{0}}V_{W,i}(r_{U}{+}v_{r}\Delta_{0}){+} (1{-}e^{-\Lambda'\Delta_{0}})
	    V_{C,i}(r_{U}{+}v_{r}\Delta_{0})\big]$,
        		\State $O_{i{+}1}(r_{U})\gets v_r^*$, where $v_r^*$ is the $\argmin$.
        		\State $\mathcal E_{W,i{+}1}(r_{U}){\gets}\epsilon^{*}(r_{U};v_r^*){+}e^{-\Lambda'\Delta_{0}}\mathcal E_{W,i}(r_{U}{+}v_r^*\Delta_{0}){+}(1{-}e^{-\Lambda'\Delta_{0}})
		\mathcal E_{C,i}(r_{U}{+}v_r^*\Delta_{0})$.
        	\EndFor
	\For{each $r_{U}{\in}[0,a]$}\Comment{Outer optimization in communication states}
        	\For{each $r{\in}[0,a]$, $\psi{\in}[0,2\pi)$ ($\mathbf s=(r_U,r,\psi)$)} \Comment{Outer optimization in communication states}
        	    \State $\hat V(\mathbf{s}){\gets}\min\Big\{\underbrace{\frac{L}{R_{GB}(r)}{+}V_{W,i}({r}_{U})}_{\xi=0},
	    \underbrace{
	    \underset{\hat{r}_{U}{\in}[0,a]}{\min}\ell_{\nu}^{*}(\mathbf{s};\hat{r}_{U}){+}V_{W,i}(\hat{r}_{U})}_{\xi=1}\Big\}$
	    \Comment{Value function given GN position}
             \State $U_{i{+}1}(\mathbf{s})\gets (\xi^*,\hat r_U^*)$, where $(\xi^*,\hat r_U^*)$ is the $\argmin$ ($\hat r_U^*=r_U$ if $\xi^*=0$).
        		\State $\hat {\mathcal E}(\mathbf{s}){\gets}\xi^*\cdot\epsilon^{*}(\mathbf{s};\hat r_U^*){+}\mathcal E_{W,i}(\hat r_U^*)$.
		\Comment{Total excess cost given GN pos., optimized over scheduling/trajectory}
        	\EndFor
	\State $V_{C,i{+}1}(r_{U}){\gets}\int_{0}^{2\pi}\frac{1}{2\pi}\int_{0}^{a}\frac{2r}{a^{2}}
	\hat V(r_U,r,\psi)\mathrm{d}r\mathrm{d}\psi'$
	\Comment{Value function in comm states, averaged over GN position}
	\State ${\mathcal E}_{C,i{+}1}(r_{U}){\gets}\int_{0}^{2\pi}\frac{1}{2\pi}\int_{0}^{a}\frac{2r}{a^{2}}
	\hat {\mathcal E}(r_U,r,\psi)\mathrm{d}r\mathrm{d}\psi'$
	\Comment{Excess energy cost in comm states, averaged over GN position}
	\EndFor
        	\State ${\forall}r_U\in[0,a]$ and $X\in\{W,C\}$, calculate $\delta_X^{(V)}(r_U){=}V_{X,i{+}1}(r_U){-}V_{X,i}(r_U)$
	and $\delta_X^{(\mathcal E)}(r_U){=}\mathcal E_{X,i{+}1}(r_U){-}\mathcal E_{X,i}(r_U)$;
	 $i{\gets}i{+}1$.
        \Until{$\max_{r_U,X}\delta_X^{V}(r_U){-}\min_{r_U,X}\delta_X^{V}(r_U){<}\delta$ and
        $\max_{r_U,X}\delta_X^{\mathcal E}(r_U){-}\min_{r_U,X}\delta_X^{\mathcal E}(r_U){<}\delta$.} \Comment{Termination condition}\\
    \Return $g(\nu){\approx}\delta_W^{(V)}(0)/\pi_{\mathrm{comm}}$,
    $\bar{\mathcal E}{\approx}\delta_W^{(\mathcal E)}(0)$.\Comment{dual cost and average excess energy cost}\\
     \ \ \ \ \ \ \ \ $V_{\cdot,0}^{next}(\cdot){=}V_{\cdot,i}(\cdot){-}V_{W,i}(0)$, $\mathcal E_{\cdot,0}^{next}(\cdot){=}\mathcal E_{\cdot,i}(\cdot){-}\mathcal E_{W,i}(0)$.
    \Comment{Relative values (next VITER initialization)}\\
     \ \ \ \ \ \ \ \ $O^{*}(\cdot){=}O_{i}(\cdot)$, $U^{*}(\cdot){=}U_{i}(\cdot)$.
     \Comment{Optimal waiting and communication policies}
    \end{algorithmic}
\end{algorithm}
\begin{algorithm} [t]
\caption{Projected Sub-gradient Ascent (PSGA)}\label{A2}
    \begin{algorithmic}[1]
    \scriptsize
    \State \textbf{Initialization}: $k=0$; dual variable $\nu{\geq}0$; step-size $\{\rho_{k}{=}\frac{\rho_{0}}{k{+}1},k{\geq}0\}$; 
    $V_{\cdot,0}(\cdot){=}\mathcal E_{\cdot,0}(\cdot)\equiv 0$.
    \Repeat 
    	\State  $(O^{*},U^{*},g,\bar{\mathcal E},V_{\cdot,0},\mathcal E_{\cdot,0})\gets\mathrm{VITER}(\nu,V_{\cdot,0},\mathcal E_{\cdot,0})$ via Alg.~\ref{A1}.
    		\State Update $\nu\gets\max\left\{\nu{+}\rho_{k}\bar{\mathcal E},0\right\}$; $k{\gets}k{+}1$. \Comment{Dual variable update}
    \Until{$\bar{\mathcal E}{<}\epsilon_{PF}$; $\nu|\bar{\mathcal E}|{<}\epsilon_{CS}$} \Comment{Check KKT optimality conditions}
    \State \textbf{return:} optimal outer policy $(O^{*},U^{*})$.
    \end{algorithmic}
\end{algorithm}
\begin{algorithm} [t]
\caption{HCSO Algorithm}\label{A3}
    \begin{algorithmic}[1]
        \scriptsize
    	\State Randomly initialize $N$ particles $(\mathbf p,\mathbf v)_{1:N}$: $\mathbf p_i$ is a sequence of way-points,  $\mathbf{v}_{i}$ a sequence of UAV speeds.
    	\While{$M \leq M_{\mathrm{max}}$}
    		\State Obtain $M$-segment trajectory: $(\mathbf{p}^{*},\mathbf{v}^{*}){=}\mathrm{CSO}(\mathbf{p}_{1:N},\mathbf{v}_{1:N},N,M)$ (see \cite{CSO}). \Comment{CSO call}
    		\State Increase $M{\gets}2M$; interpolate to form reference trajectory: $(\tilde{\mathbf{p}},\tilde{\mathbf{v}}){=}\mathrm{interp}(\mathbf{p}^{*},\mathbf{v}^{*},M)$. \Comment{Increase resolution via interpolation}
    		\State Reduce swarm size $N{\gets}N{-}N_{\mathrm{red}}$.
    		\For{$n{=}1,2,{\dots},N$}\Comment{Generate $N$ particles randomly}
              \State New way-point particle $\mathbf p_{n}$ with $m$th way-point $\mathbf x_m=\tilde{\mathbf{x}}_{m}{+}(\chi_{m},\zeta_{m})$
              and $\mathbf x_M=\hat r_U\frac{\mathbf{x}_{M{-}1}}{\Vert\mathbf{x}_{M{-}1}\Vert_2}$.
 \Comment{Way-point perturbation}
              \State New velocity particle $\mathbf v_{n}$ with $m$th velocity $v_m=[\tilde{v}_{m}{+}\varkappa_{m}]^{[V_{\mathrm{low}},V_{\mathrm{max}}]}$. \Comment{Velocity perturbation}
    		\EndFor
    	\EndWhile
    \end{algorithmic}
\end{algorithm}
\vspace{-2mm}

\section{Trajectory Design via Hierarchical Competitive Swarm Optimization}\label{S4}
In this section, we design the UAV trajectory during the D\&F protocol. To solve \eqref{eq:EllMin}, we propose a CSO scheme \cite{CSO} defining a \emph{meta-heuristic UAV trajectory}. First, as done also with SCA approaches \cite{SCA,CSCA-ADMM,EnergyEfficientUAVs}, we simplify the continuous UAV trajectory into a finite sequence of way-points connected by straight lines at constant velocity. However, a direct application of CSO to high-resolution trajectory design suffers from poor convergence due to exponentially large solution spaces \cite{HighDimensionality}. We address this weakness by proposing a Hierarchical variant of CSO (HCSO), wherein a sequence of problems is solved: initially, CSO produces a low-resolution trajectory; the optimized trajectory is then interpolated to create a higher-resolution one, then further optimized with CSO. The process repeats until a target resolution is achieved.

Let $\mathbf x_0=(r_U,0)$ be the initial UAV position and $\mathbf{x}_G{\triangleq}(r\cos{\psi},r\sin{\psi})$ be the request position (in this section, expressed as Cartesian coordinates), corresponding to the communication state $\mathbf{s}{=}(r_{U},r,\psi){\in}\mathcal{S}_{\mathrm{comm}}$. Given a target end radius position $\hat{r}_{U}$ (the outer action), we encode the UAV trajectory as a sequence of $M$ way-points $\mathbf{x}_{m}{=}(x_{m},y_{m}),m=1,\dots,M$, ending at $\mathbf{x}_{M}$ at radius $\hat r_U$, and velocities $v_{m}{\in}[V_{\mathrm{low}},V_{\mathrm{max}}]$ used to traverse each straight trajectory segment $\Psi_{m}{\triangleq}\mathbf{x}_{m}{-}\mathbf{x}_{m{-}1}$. The first and second $\frac{M}{2}$ segments correspond to the two phases of the D\&F protocol. Here, the minimum velocity $V_{\mathrm{low}}{\ll}V_{\mathrm{max}}$  ensures well-defined segment durations; the sequences of way-points $\mathbf{p}{\triangleq}[\mathbf{x}_{1},{\dots},\mathbf{x}_{M}]$ and velocities $\mathbf{v}{\triangleq}[v_{1},{\dots},v_{M}]$ are the optimization variables. Since the number of bits communicated \eqref{eq:PLConst1} during each trajectory segment, coupled with our throughput-maximizing rate adaptation scheme, cannot be computed in closed-form, we approximate them numerically. Specifically, between subsequent way-points $\mathbf{x}_{m-1}$ and $\mathbf{x}_{m}$ traversed with velocity $v_{m}$, we generate a sequence of $n_{\mathrm{res}}$ evenly-spaced points with sufficiently high resolution; \hlt{letting $\{R_{k}^{\mathrm{new}}\}_{k{=}1}^{n_{\mathrm{res}}}$ be the expected throughput at each point,
computed via \eqref{TBar} and Prop. \ref{P1}, the number of bits communicated along the $m$th segment is approximated as $F_{m}{\triangleq}\frac{\Vert\Psi_{m}\Vert_2}{v_m}\frac{1}{n_{\mathrm{res}}}\sum_{k{=}1}^{n_{\mathrm{res}}}R_{k}^{\mathrm{new}}$, where $\frac{\Vert\Psi_{m}\Vert_{2}}{v_{m}}$ is the time taken to traverse it.} Thus, \eqref{eq:EllMin} becomes
\begin{align}
    &(\mathbf{P.0})\;\;\; \underset{\mathbf p,\mathbf v{\in}[V_{\mathrm{low}},V_{\mathrm{max}}]^M}{\mathrm{min}} \;\,  \sum_{m=1}^{M} \frac{\Vert \Psi_m \Vert_2}{v_m}\Big(1-2\alpha+\alpha \frac{P_{\mathrm{mob}}(v_m)}{P_{\max}}\Big) \label{eq:HeurMin}\\
    &\mathrm{s.t.}\; h_{i}(\mathbf p ,\mathbf v) \triangleq L - \sum_{m=\frac{M}{2}i+1}^{\frac{M}{2}(i+1)} F_{m} \leq 0, \;\; i = 0\text{ and }1,\label{eq:NewPLC}\tag{$\tilde{\text{C}}$}
    \Vert\mathbf x_M\Vert_2 = \hat{r}_U,
\end{align}
where \ref{eq:NewPLC} enforce the data payload and end radius constraints. 
To solve $(\mathbf{P.0})$ with CSO, we first convert it into an unconstrained one,  by penalizing constraint violations with a particular solution: 1) if the UAV does not decode (or forward) its data payload by the end of either phase, then it flies along the circumference of a circle (radius $r_{\mathrm{min}}{>}0$, small) around the current position with its power-minimizing velocity ($v_{P_{\mathrm{min}}}${=}22 m/s \cite{SCA}) until the transmission/reception is completed; and 2) we enforce the end radius constraint
by projecting the penultimate way-point $\mathbf{x}_{M{-}1}$ to the circle at radius $\hat{r}_{U}$, i.e.
$\mathbf x_M=\hat r_U\mathbf{x}_{M{-}1}/\Vert\mathbf{x}_{M{-}1}\Vert_2$.\footnote{We let $\frac{\mathbf{x}}{\Vert\mathbf{x}\Vert_{2}}{=}(1,0)$ for a point in the origin, $\mathbf{x}{=}(0,0)$.}
This yields the penalized objective function
\begin{align*}
    &\hat{f}(\mathbf{p},\mathbf{v}){\triangleq}\sum_{m{=}1}^{M}\frac{\Vert\Psi_{m}\Vert_{2}}{v_{m}}\Big(1-2\alpha+\alpha \frac{P_{\mathrm{mob}}(v_m)}{P_{\max}}\Big){+}(1-2\alpha)(\hat{t}_{P,0}{+}\hat{t}_{P,1}){+}\alpha \frac{\hat{E}_{P,0}{+}\hat{E}_{P,1}}{P_{\max}};\nonumber\\
    &\hat{t}_{P,0}{\triangleq}\frac{\max\{h_{0}(\mathbf{p},\mathbf{v}),0\}}{\bar{R}_{GU}(\Vert\mathbf{x}_{M/2}-\mathbf{x}_{G}\Vert_{2})};\ 
    \hat{t}_{P,1}{\triangleq}\frac{\max\{h_{1}(\mathbf{p},\mathbf{v}),0\}}{\bar{R}_{UB}(\Vert\mathbf{x}_{M}\Vert_{2})};\ \hat{E}_{P,i}{\triangleq}
    P_{\min}\hat{t}_{P,i},
    \ \mathbf x_M=\hat r_U\frac{\mathbf{x}_{M{-}1}}{\Vert\mathbf{x}_{M{-}1}\Vert_2},
\end{align*}
where $\hat{t}_{P,i}$ and $\hat{E}_{P,i}$ are the time and energy penalties involved in finishing the data communication during the decode and forward phases ($i{=}0\text{ and }1)$. 
In particular, $\hat{t}_{P,i}$ equals the remaining payload
$\max\{h_{i}(\mathbf{p},\mathbf{v}),0\}$, divided by the
corresponding throughput at the terminal position ($\bar R_{GU}$ for the decode phase and $\bar R_{UB}$ for the forward phase). Hence, ($\mathbf{P.0}$) becomes $\underset{\mathbf{p},\mathbf{v}}{\mathrm{min}}\hat{f}(\mathbf{p},\mathbf{v})$. To solve this problem, we employ the HCSO algorithm, outlined in Alg.~\ref{A3} and discussed next.

We initialize $N$ way-point particles $\mathbf{p}_{1:N}{\triangleq}\mathbf{p}_{1},{\dots},\mathbf{p}_{N}$ and $N$ UAV velocity particles $\mathbf{v}_{1:N}\triangleq\mathbf{v}_{1},\dots,\mathbf{v}_{N}$ (line 1). The core of the algorithm is CSO (line 3), detailed in \cite{CSO}: essentially, during the $k$th iteration within CSO, the  $N$ particles are randomly grouped into $\frac{N}{2}$ pairwise competitions. For both members of a pair, $\hat{f}(\mathbf{p},\mathbf{v})$ is calculated; the winner of the competition is passed onto the $(k{+}1)$th iteration, while the loser is modified by learning from the winner, as detailed by the update equations in \cite{CSO};
after repeating these pair-wise competitions, the CSO algorithm outputs a winning trajectory $(\mathbf{p}^*,\mathbf{v}^*)$. \label{dilemma}\hlt{However, a direct application of CSO alone suffers from a complexity-accuracy dilemma: high-resolution trajectories are slow to converge, while low-resolution ones give rise to poor solutions that fail to capture fine-grained variations in the trajectory way-points and velocities. To overcome this limitation, we embed CSO  within a hierarchical wrapper: starting from a low-resolution trajectory optimized via CSO,}
after each CSO iteration (line 3), the resulting trajectory is interpolated to form a reference higher-resolution trajectory of $M{\gets}2M$ way-points (line 4). The new population size is then reduced, $N{\gets}N{-}N_{\mathrm{red}}$, to lower the computational burden of CSO (line 5), and a new set of $N$ particles is generated randomly. \hlt{To preserve the quality of the previous lower-resolution trajectory solution,} the $m$th way-point of each new particle is generated by injecting zero-mean Gaussian noise $\chi_{m},\zeta_{m}{\sim}\mathcal{N}\left(0,\sigma_{m,X}^2\right)$ (line 7) around the reference trajectory; similarly, the UAV velocity is
generated by injecting Gaussian noise $\varkappa_{m}{\sim}\mathcal{N}(0,\sigma_V^2)$ (line 8), followed by projection onto the feasible set ($[\cdot]^{[V_{\mathrm{low}},V_{\mathrm{max}}]}$). Here, the way-point variance  $\sigma_{m,X}^2=\varsigma(\Vert\tilde{\mathbf{x}}_{m{+}1}{-}\tilde{\mathbf{x}}_{m}\Vert^{2}{+}\Vert\tilde{\mathbf{x}}_{m{-}1}{-}\tilde{\mathbf{x}}_{m}\Vert^2)$, with scaling factor $\varsigma{>}0$, is determined by the spread between neighboring reference trajectory way-points. This choice accounts for the empirical observation that in areas with clustered UAV way-points, the objective function $\hat{f}(\mathbf{p},\mathbf{v})$ is sensitive to large variations. The speed variance $\sigma_V^2=\varepsilon(V_{\mathrm{max}}{-}V_{\mathrm{low}})^{2}$, with scaling factor $\varepsilon{>}0$, reflects the observation that the UAV velocities exhibit faster convergence with CSO than the trajectory way-points and less sensitivity to random initialization. These steps in Alg.~\ref{A3} continue until the desired trajectory resolution is reached. 
\begin{figure} [t]
    \centering
    \includegraphics[width=1.0\linewidth]{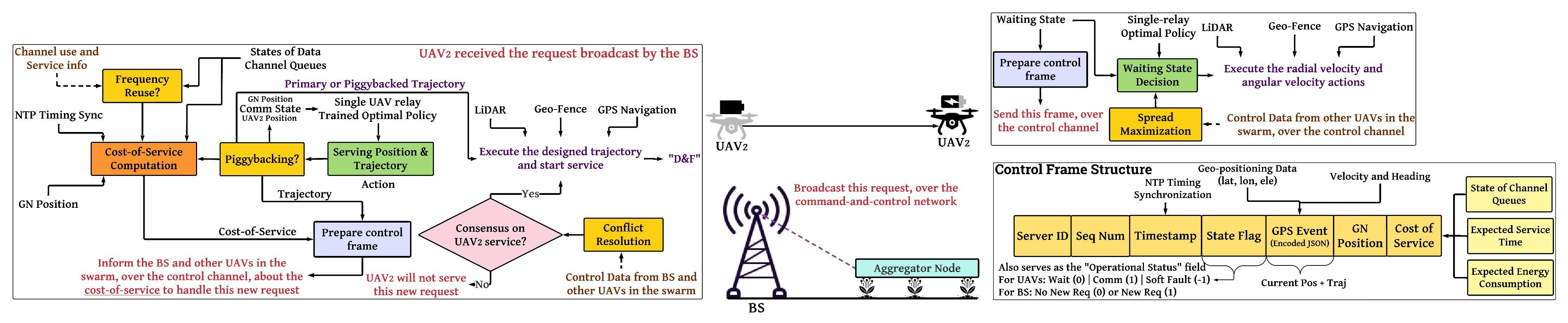}
    \vspace{-8mm}
    \caption{\hlt{An illustration outlining the sequence of operations under MAESTRO-X that occur at each UAV.}}
    \label{F5}
\end{figure}
\vspace{-4mm}

\section{MAESTRO-X: An Extension to UAV Swarms}\label{S5}
In this section, we extend MAESTRO to swarms of $N_{U}$ UAV-relays. This eXtension, termed MAESTRO-X, augments the multiscale optimal policy obtained via SMDP value iteration. Depicting an example scenario of serving data traffic generated by an aggregation of soil sensors in precision agriculture, Fig.~\ref{F5} illustrates its control flow. 
\hlt{MAESTRO-X is enabled by replicating the optimal single-agent policy of the SMDP in Sec.~\ref{S3} across the swarm and employing additional enhancements including \emph{spread maximization}, \emph{consensus-driven conflict resolution} with queuing dynamics, \emph{piggybacking}, and \emph{frequency reuse}. These mechanisms\footnote{Due to space constraints, we keep our discussions on these multi-agent mechanisms brief. For more details on their implementation, please refer to our source code on \href{https://github.com/bharathkeshavamurthy/MAESTRO-X.git}{GitHub} \cite{MAESTRO-X}.} are implemented using a fully-connected distributed mesh network overlaid on the BS and UAVs,
that enables periodic exchanges of command-and-control messages, as depicted in Fig.~\ref{F5}.}

\noindent{\textbf{Spread Maximization}}: Note that the inner action of MAESTRO's optimal waiting policy is symmetric in relation to clockwise and counter-clockwise angular UAV movements. For multiple UAVs, we leverage this symmetry to proactively position idle UAVs for potential new relay requests. Specifically, each UAV in the waiting state moves either clockwise or counter-clockwise (with angular velocity given by \eqref{eq:MinLWP}), so as to maximize its angular distance from the nearest UAV in the waiting state, in an attempt to spread out and more readily serve future requests. To this end, UAV $i$ parses the state flag as 0 and GPS event fields in its control frame (see Fig.~\ref{F5}). By monitoring the control frames received from other UAVs, it constructs a local peer list $\mathcal{L}$ of other waiting state UAVs, and determines its closest peer (in the angular dimension) $j^{*}{=}\argmin_{j{\in}\mathcal{L}}|\theta_{i}{-}\theta_{j}|$, where $\theta_{j}$ is the current angular coordinate of UAV $j$. UAV $i$ then executes the angular motion away from UAV $j^*$, until new control frames (containing updated positions) are received from its peers (at the end of the synchronized reporting period) or upon receiving a new GN transmission request, at which time it transitions to the communication state.

\noindent{\textbf{\hlt{Consensus-driven Conflict Resolution}}}: \hlt{In our single-UAV formulation (Sec.~\ref{S3}), the scheduling action was determined by comparing the Lagrangian costs of direct-BS transmission to that of relayed UAV service. To extend scheduling decisions to UAV swarms---including queueing dynamics, as well as simultaneous multi-user service via piggybacking at the UAVs and frequency reuse (both described later in this section)---the augmented scheduling decision must now 1) resolve conflicts among the BS and UAVs as to whom should serve a new GN request; 2) facilitate a consensus on the best node to serve the GN; 3) account for queueing delays experienced at each potential server node while waiting for data channels to become available. Similarly to the single-UAV setting, this augmentation is driven by a cost-of-service metric computed at the BS and at each UAV. The new metric consists of several modifications to the original delay-energy cost trade-off computed in the single-UAV setting. For new requests served directly by the BS, the new metric equals the original delay metric, plus an estimate of the time needed for a data channel to become available (and considers the frequency reuse mechanism to be described). This time can be estimated based on the time needed to complete the requests currently served at the BS, and the time needed to complete those already queued. Thus, for a new GN request at $(r,\theta)$, the augmented cost metric associated with direct-BS transmission is $\frac{L}{\bar{R}_{GB}(r)}{+}t_{\mathrm{BS}}$, where the first term accounts for the transmission time, whereas $t_{\mathrm{BS}}$ is the additional waiting time.}

\hlt{Meanwhile, for new requests served by UAV $i$ at radius $r_{U|i}$, GN request radius $r$, and angle between them $\psi_{U|i}$, i.e., state $\mathbf s_i = (r_{U|i}, r, \psi_{U|i})$, with target end radius $\hat r_{U|i}$, the augmented cost metric is given by $\tilde{\ell}_{\nu}^* (\mathbf s_i; 1, \hat r_{U|i}) + t_{\mathrm{U|i}}$. The first term, $\tilde{\ell}_{\nu}^* (\mathbf s_i; 1, \hat r_{U|i})$, is the Lagrangian cost metric, modified to account for the piggybacking mechanism (described later in this section), wherein the UAV follows a collated trajectory to handle the new request while serving previous requests; the second term, $t_{\mathrm{U|i}}$, is an estimate of the time needed for a data channel to become available (and considering the frequency reuse mechanism). Upon calculating these cost-of-service metrics for the BS and the UAVs, the network arrives at a consensus on the best node to serve the new request, i.e., if $\frac{L}{\bar{R}_{GB} (r)}{+}t_{\mathrm{BS}}{\leq}\tilde{\ell}_{\nu}^* (\mathbf s_i; 1, \hat r_{U|i}){+}t_{\mathrm{U|i}}$,${\forall}i{\in}\{1,2,{\dots},N_U\}$, then the BS serves the request; otherwise, the request is relayed through the UAV $i^*{=}\arg\min_{i{\in}\{1,2,{\dots},N_U\}}\tilde{\ell}_{\nu}^* (\mathbf s_i; 1, \hat r_{U|i}) + t_{\mathrm{U|i}}$.}

\phantomsection\label{freq_reuse_label}
\noindent{\hlt{\textbf{Frequency Reuse}}}: \hlt{To improve the spectrum utilization efficiency, we propose a frequency reuse mechanism, allowing multiple serving nodes (the BS and UAVs) to share the same data channel simultaneously when serving their respective GN requests. When direct-BS transmission is used to serve a new GN request, a single data channel assignment occurs at the start of direct transmission. When the new request is instead served using a D\&F UAV relay, two distinct data channel assignments occur: one each for the decode and forward phases of the UAV. In essence, reuse of an occupied data channel is permitted on the condition that the received SNRs of nodes sharing the data channel degrade no more than an acceptable pre-specified threshold permits. Moreover, to make operations of the frequency reuse mechanism more amenable to our problem, which includes UAVs following time-varying trajectories, we equivalently describe this SNR degradation threshold by instead using a minimum distance threshold $d_{\mathrm{th}}$.}

\hlt{The frequency reuse mechanism proceeds in the same way, regardless of whether the data channel assignment under consideration is for a GN using direct-BS transmission, a GN sending its data to a UAV (decode phase), or a UAV relaying its data payload to the BS (forward phase). To formalize, let $k{\in}\{1,2,{\dots},N_C\}$ be the data channel under consideration for reuse; let node $i$ be the new transmitter (either a GN beginning its uplink transmission or a UAV beginning its forward phase) determining whether reuse of data channel $k$ is possible; let node $j$ be the intended receiver of the transmission originating from node $i$; let $\mathcal T (k)$ be the set of active transmitters already using data channel $k$ to serve their requests, i.e., a GN transmitting to a BS or UAV, or a UAV transmitting to the BS during its forward phase; let $\mathcal R (k)$ be the set of active receivers already using data channel $k$, i.e., a UAV receiving an uplink transmission from a GN during the decode phase, the BS receiving an uplink transmission directly from a GN, or the BS receiving the data payload from a UAV during the forward phase. For data channel $k$ to be deemed acceptable for reuse, the following two conditions must both be met:
\begin{align}
    &(\textbf{FR.1}) \;\;\; d_{\ell,j} \geq d_{\mathrm{th}}, \; \forall \ell \in \mathcal T (k), \\
    &(\textbf{FR.2}) \;\;\; d_{i,\ell} \geq d_{\mathrm{th}}, \; \forall \ell \in \mathcal R (k),
\end{align}
where $d_{i',j'}$ is the Euclidean distance between any transmitter $i'$ and receiver $j'$. From the above equations, $(\textbf{FR.1})$ ensures that the distances between the intended receiver and all currently active transmitters are above the minimum distance threshold $d_{\mathrm{th}}$, at all times during the execution of the UAVs' trajectories. Likewise, $(\textbf{FR.2})$ ensures that distances between the new transmitter and all currently active receivers are above the minimum distance threshold $d_{\mathrm{th}}$. Effectively, satisfying conditions $(\textbf{FR.1})$ and $(\textbf{FR.2})$ simultaneously ensure that no received SNR experiences a degradation beyond a pre-specified limit, and hence data channel $k$ is acceptable for reuse. Next, given its re-usability, the wait time for a channel to become available is estimated by modeling queuing dynamics, choosing the channel with the smallest wait time for service. Also, note that, once a channel is chosen with reuse, since the throughput experienced by the UAV during service degrades due to the added interference from other transmitters using the same channel, the UAV might not be able to complete its decode or forward phases using the optimal trajectory: the UAV then flies along the circumference of a circle ($r_{\min}{>}0$) around the phase-specific final way-point with its power-minimizing velocity (22 m/s) to complete the phase; additionally, we evaluate the service in this case using the same time and energy penalties discussed in Sec.~\ref{S4}.}

\phantomsection\label{pb_label}
\noindent{\hlt{\textbf{Piggybacking}}}: \hlt{To facilitate simultaneous multi-user service at the UAVs, we incorporate a piggybacking mechanism (in the cost-of-service computation of the consensus-driven conflict resolution process), wherein a UAV follows a collated trajectory to accommodate new GN uplink requests while serving previous requests. Recalling from the description of conflict resolution, for a new request served through UAV $i$, we consider the state $\mathbf s_i = \left( r_{U|i}, r, \psi_{U|i} \right)$, with target end radius $\hat r_{U|i}$, and modified Lagrangian cost metric $\tilde{\ell}_{\nu}^{*} (\mathbf s_i; 1, \hat r_{U|i} )$. If UAV $i$ is currently not serving any other request, this modified cost metric simplifies to $\tilde{\ell}_{\nu}^* (\mathbf s_i; 1, \hat r_{U|i}) {=} \ell_{\nu}^* (\mathbf s_i; 1, \hat r_{U|i})$, i.e., the original Lagrangian cost metric computed for the single UAV. On the other hand, if the UAV is currently serving other requests, the UAV computes the cost metric to serve the new request by \emph{piggybacking} it, i.e., serving it simultaneously with its current requests on a different data channel. In this case, the modified cost metric becomes $\tilde{\ell}_{\nu}^* (\mathbf s_i ; 1, \hat r_{U|i} ) {=} \ell_{\nu}^{(\mathrm{pg})} (\mathbf s_i; 1, \hat r_{U|i})$, where $\ell_{\nu}^{(\mathrm{pg})} (\mathbf s_i; 1, \hat r_{U|i})$ is defined to encapsulate modifications to the cost-of-service metric corresponding to the amount of data payload of the new request that has been either decoded or forwarded (or both) during the execution of the current trajectory (serving the UAV's previous requests). Note that the energy expended by the UAV serving its current trajectory while piggybacking the new request is not considered in the cost computed for this new request, since the energy cost has already been accounted for in the execution of the current trajectory; instead, we consider only the delays experienced by the piggybacked GN during its associated cost computation.}
\vspace{-4mm}

\section{Simulation Setup and Evaluations}\label{S6}
\hlt{Unless otherwise stated, we use the parameter values in Table~\ref{T_params}. To solve \eqref{eq:PolDecomp} via Algorithms \ref{A1}--\ref{A3}, we discretize the SMDP state and action spaces (with 25 equally-spaced radii levels and 25 radial velocity waiting actions) and apply linearly-interpolated value iteration (see implementation details documented in \cite{MAESTRO-X}). 
 Furthermore,  we chose $\Delta_{0}=1$s.}
\begin{table}
\hlt{\begin{center}
\scriptsize
    \begin{tabular}{|*{7}{c|}}
    \hline
    \thead{\textbf{Notation}} & \thead{\textbf{Description}} & \thead{\textbf{Simulation Value}} & & \thead{\textbf{Notation}} & \thead{\textbf{Description}} & \thead{\textbf{Simulation Value}}\\
    \hline
    $N_G$ & Number of GNs & 30 & & $a$ & Cell radius & 1 km \\
    \hline
    $L$ & Data payload & 10 Mbits & & $W$ & System BW & 20 MHz \\
    \hline
    $N_C$ & Number of data channels & 4 & & $B$ & Data channel BW & 5 MHz \\
    \hline
    $\kappa$ & NLoS attenuation constant & 0.2 & &  & SNR referenced at 1 m & 40 dB \\
    \hline
    $(\alpha, \tilde{\alpha})$ & LoS/NLoS pathloss exponents & (2,2.8) & &  & UAV mobility power consumption & 
    Eq. \eqref{eq:Power}, params. of \cite{SCA}
    \\
    \hline
    $(k_1,k_2)$ & Rician $K$-factor parameters \cite{Rician} & (1,0.05) & & $(z_1,z_2)$ & LoS probability parameters \cite{OptimalAltitude} & (9.61,0.16) \\
    \hline
    $H_U$ / $H_B$ & UAV / BS antenna height & 200 m / 80 m & & $V_{\max}$ & Max. UAV speed & 55 m/s \\
    \hline
    & Control frame reporting period & 10 ms & &
    & SINR degradation threshold & 5 dB \\
    \hline
    \end{tabular}
    \vspace{-2mm}
    \caption{\hlt{The system simulation parameters (unless otherwise stated).}}\label{T_params}
    \vspace{-4mm}
\end{center}}
\end{table}

\label{Validation}
\noindent{\hlt{\underline{Validation of surrogate optimization problem \eqref{minW}}}: \hlt{First, we justify the efficacy of our alternative optimization framework that replaces the original metric $\bar{D}_{\mu}$ with the lower bound $\bar{W}_{\mu}^{(s)}$. As depicted in Table~\ref{T4}, we observe that the optimized value $\bar{W}_{\mu^*}^{(s)}$ of the alternative formulation \eqref{minW} is practically identical to the expected delay metric $\bar{D}_{\mu^*}$ of the original formulation \eqref{eq:opt_prob_orig}, across various data payload sizes ($L$) and data traffic arrival rates ($\Lambda'$). Hence, replacing $\bar{D}_{\mu}$ with its lower bound $\bar{W}_{\mu}^{(s)}$ as the optimization metric leads to  near-optimal solutions. Notably, the surrogate optimization problem \eqref{minW} is amenable to dynamic programming tools such as value iteration (see Alg.~\ref{A1}) and enables our proposed two-scale policy decomposition that drastically reduces the size of the action space in our SMDP formulation. These tools would not be directly applicable to the original formulation \eqref{eq:opt_prob_orig} that uses $\bar{D}_{\mu}$ as the optimization objective.}
\begin{table}[t]
\hlt{\begin{center}
\scriptsize
    \begin{tabular}{|*{6}{c|}}
    \hline
    \thead{Payload: $L$} &
    \thead{Arrival rate: $\Lambda'$} &
    \thead{Lower bound: $\Bar{W}_{\mu^{*}}^{(s)}$} &
    \thead{Expected Delay: $\Bar{D}_{\mu^{*}}$} &
    \thead{Direct-to-BS: $\Bar{D}_{BS}$}\\
    \hline
    1 Mbits & 1 req/min/UAV & 1.15 s & 1.15 s & 31.64s\\
    \hline
    10 Mbits & 0.2 req/min/UAV & 16.41 s & 16.41 s & 316.38 s\\
    \hline
    100 Mbits & 0.033 req/min/UAV & 82.17 s & 82.17 s & 3163.81 s\\
    \hline
    \end{tabular}
    \vspace{-2mm}
    \caption{\hlt{$P_{\mathrm{avg}}{=}$1 kW: A comparison between the lower bound $\Bar{W}_{\mu^{*}}^{(s)}$ of $\Bar{D}_{\mu^{*}}$ (Prop.~\ref{P2}) and direct-BS ($\Bar{D}_{BS}$).}}\label{T4}
    \vspace{-4mm}
\end{center}}
\end{table}}
\begin{figure} [t]
    \begin{subfigure}{0.485\linewidth}
      \centering
      \includegraphics[width=1.0\linewidth]{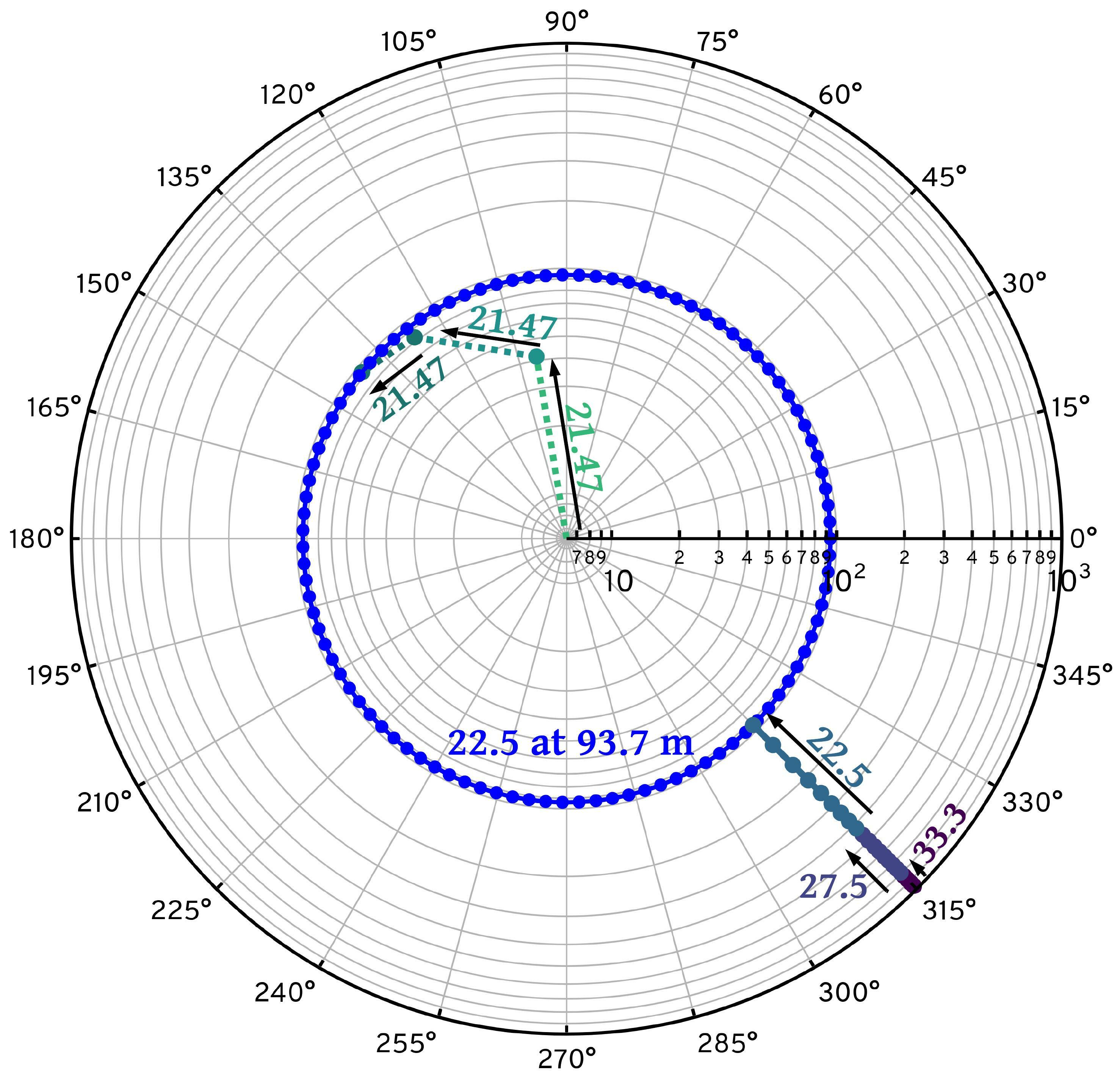}
      \caption{\hlt{Optimal Wait Policy.}}
      \label{F6}
    \end{subfigure}
    \hfill
    \begin{subfigure}{0.515\linewidth}
      \centering
      \includegraphics[width=1.0\linewidth]{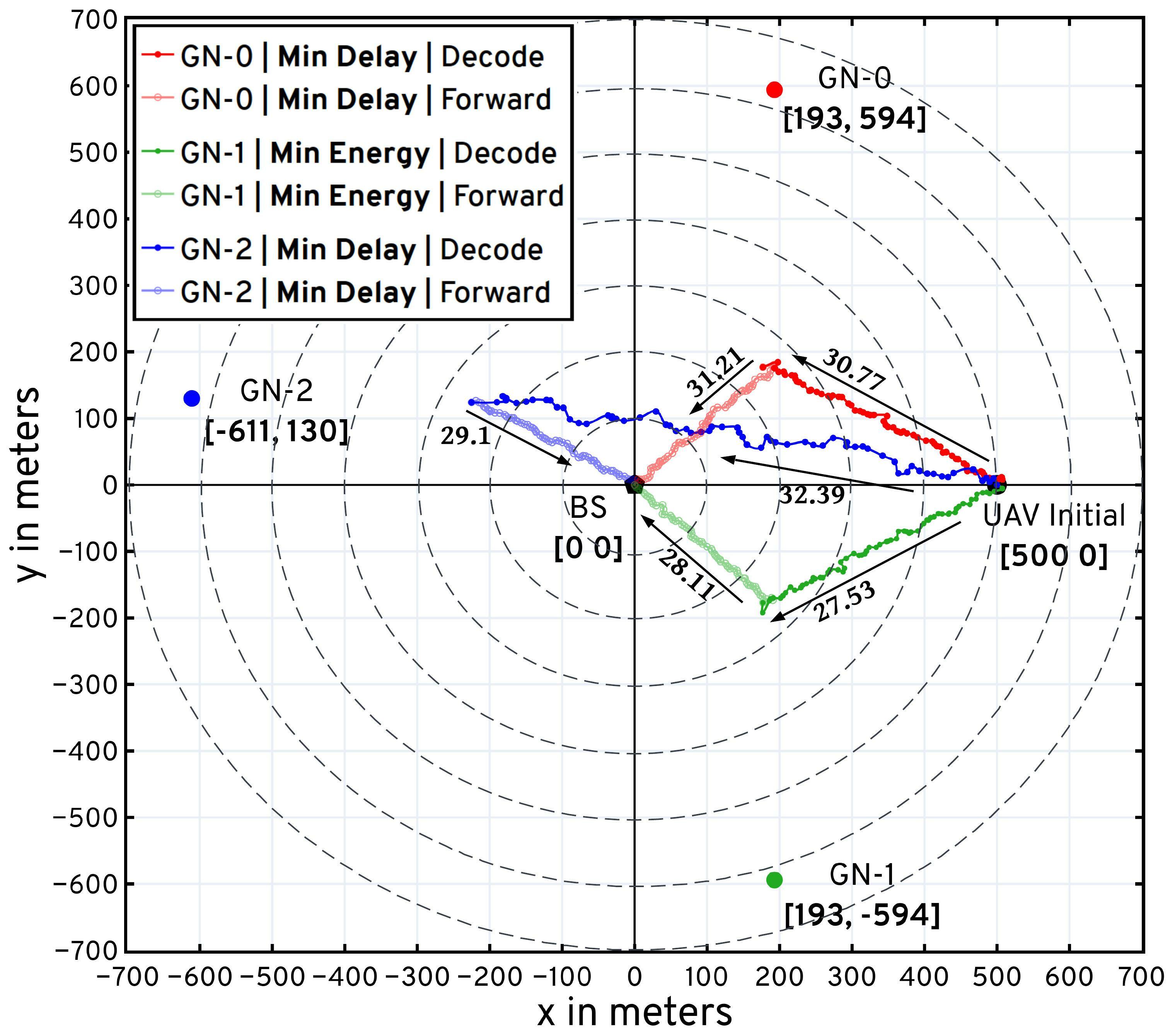}
      \caption{\hlt{Optimal D\&F trajectory.}}
      \label{F7}
    \end{subfigure}
    \caption{\hlt{$L{=}$10 Mbits, $P_{\mathrm{avg}}{=}$1.2 kW,
    $\Lambda'{=}$0.2 req/min/UAV: Optimal waiting policy (a) and optimized D\&F trajectory during a communication phase 
    (terminating above the BS) (b).
    The arrows and associated numerical values represent the direction of motion and the flying speed in m/s.}}
    \label{Fig4}
\end{figure}

\noindent{\underline{MAESTRO policy}}: We now study illustrative examples of the optimal policy (Fig. \ref{Fig4}). We note that, during the waiting phase (Fig.~\ref{F6}), the UAV moves towards a radius of $\approx$94 m; upon reaching it, it flies at power-minimizing speed (22.5 m/s) along a circle: this allows the UAV to be well-positioned for future requests (not too close to the BS, and not too far away from it), and at the same time to minimize its power consumption. Next, Fig.~\ref{F7} depicts the optimal trajectory obtained via HCSO (Algorithm \ref{A3}), for a certain configuration of GN request positions, initial and target final UAV radii (evident from the figure). \hlt{Intuitively, during the decode phase, the UAV flies towards the GN to improve the pathloss conditions; for the same reason, it moves towards the BS during the forward phase. Additionally, Fig~\ref{F7} depicts two different trajectory choices for the GNs at [193, $\pm$594] m (GN-0 and GN-1, specular to each other), one corresponding to minimum service delay and the other corresponding to minimum service energy: here, in addition to observing the angular symmetry in our formulation (see Sec.~\ref{S3}), we notice that, under the minimum delay trajectory, the UAV flies faster, to improve pathloss quicker and reduce the transmission delay; in contrast, it flies slower under the minimum energy trajectory, to save energy.
The delay-energy trade-off in trajectory design is regulated via $\alpha$, as described by \eqref{eq:EllMin}.}

\noindent{\underline{MAESTRO-X delay-power trade-off}}: We compare the delay-power trade-off of MAESTRO-X with adaptations of state-of-the-art algorithms to our setup, namely: the \emph{CIRCLE} heuristic \cite{MEC-DDPG}; a CVXPY implementation of the Successive Convex Approximation scheme (\emph{SCA}) \cite{SCA}; a CVXPY implementation of the Constrained SCA scheme with Alternating Direction Method of Multipliers (\emph{CSCA-ADMM}) \cite{CSCA-ADMM}, and a TensorFlow implementation of the Double Deep-Q Networks framework (\emph{DDQN}) \cite{DDQN}. Note that all these frameworks are optimized under their original channel and communication models detailed in the corresponding references (see Table \ref{T1} for a list of their features), while we evaluate their performance under more realistic models of dynamic traffic arrivals and A2G channels. In addition, we consider the following custom heuristics: \emph{BS-only}, in which GNs transmit directly to the BS without using UAVs; \emph{HAP-only} in which GNs transmit directly to a High Altitude Platform (HAP, height=2 km); and \emph{Static}, in which the UAVs statically hover at fixed locations.
\label{LB}
\hlt{We also compute a \emph{Lower Bound} to the delay as follows:
for a GN at radius level $r$,
it is the minimum between the delay incurred with direct-BS transmission (with throughput $\bar{R}_{GB}(r)$), and a D\&F scheme in which the UAV is on top of the GN during the decode phase (with throughput $\bar{R}_{GU}(0)$), and
on top of the BS during the forward phase (with throughput $\bar{R}_{UB}(0)$). Note that this lower bound is not attainable, since it neglects the mobility of the UAV.}
We average the results over 1000 requests.
\begin{figure} [t]
     \begin{subfigure}{0.55\linewidth}
         \centering
         \includegraphics[width=1.0\linewidth]{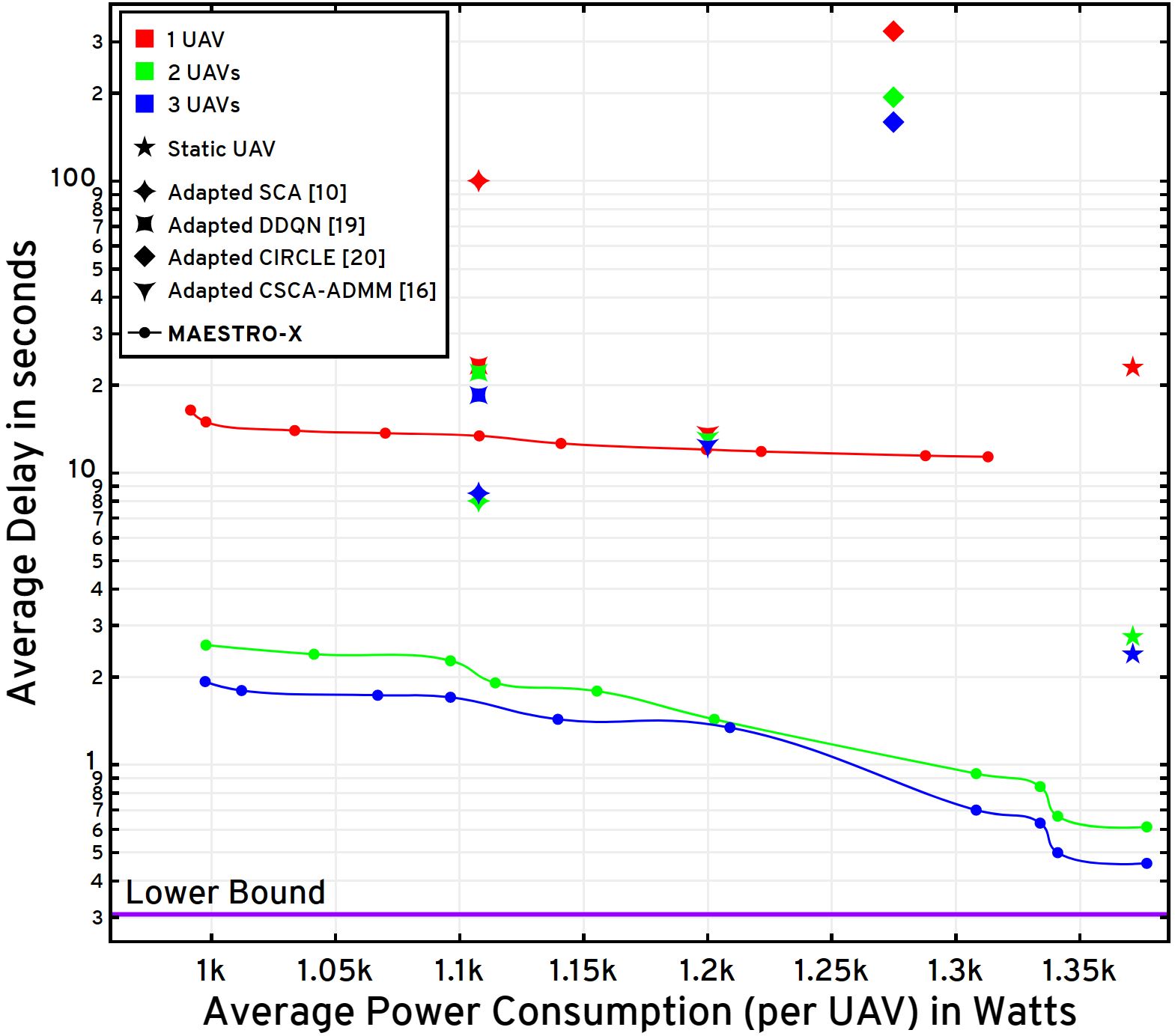}
         \caption{\hlt{Delay-Power Trade-off.}}
         \label{F8}
     \end{subfigure}
     \hfill
     \begin{subfigure}{0.433\linewidth}
         \centering
         \includegraphics[width=1.0\linewidth]{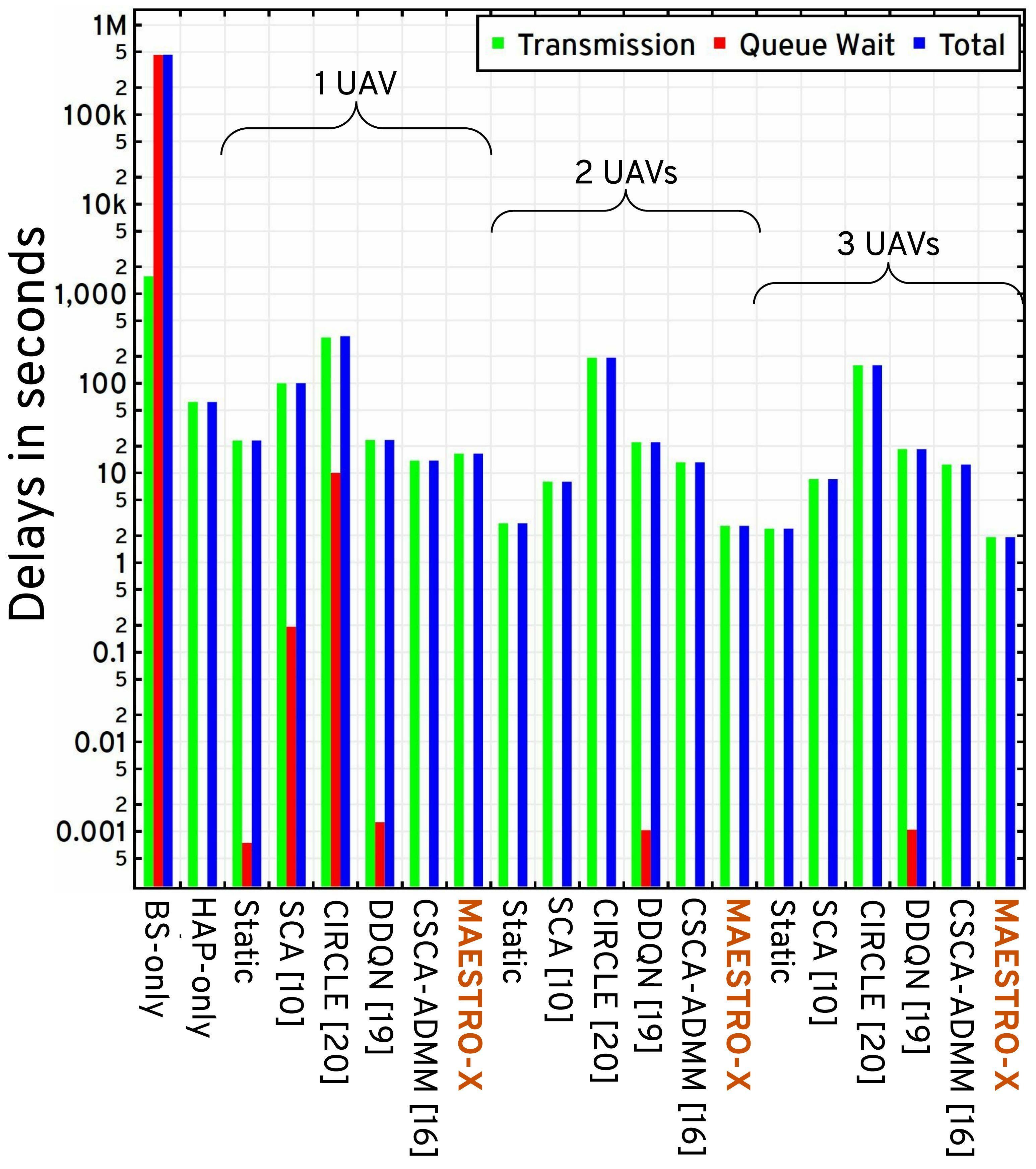}
         \caption{\hlt{Delay chart.}}
         \label{F9}
     \end{subfigure}
     \vspace{-2mm}
     \caption{\hlt{$L{=}10$ Mbits, $\Lambda'{=}$0.2 req/min/UAV: Delay-power trade-off (a) and delay charts (b) for MAESTRO-X, state-of-the-art algorithms, and custom heuristics. In (b), MAESTRO-X is evaluated under $P_{\mathrm{avg}}=$1 kW.}}
     \label{F8F9}
\end{figure}

In Fig.~\ref{F8}, we plot the delay-power trade-off under low congestion ($\Lambda'{=}$0.2 req/min/UAV).
\hlt{Remarkably, MAESTRO-X allows to regulate the delay-power trade-off, whereas the other schemes do not.
Across such trade-off, it outperforms all other schemes.}
\hlt{Specifically,
exploiting the mobility and maneuverability of the UAVs via optimized trajectories demonstrate lower service delays compared to static UAV deployments: for instance, a single UAV optimized via MAESTRO under 1 kW power constraint delivers the data payload 29\% faster than a static UAV, while using 27\% less power. 
Notably, under the same power consumption as the competitors, a single UAV optimized with MAESTRO achieves 38\% lower delay than 3 UAV relays  under DDQN \cite{DDQN}, and 13$\times$ faster service times than the CIRCLE heuristic with 3 UAVs \cite{MEC-DDPG}}. 
Adding UAVs significantly improves the performance of MAESTRO-X: with 3 UAVs MAESTRO-X delivers the payloads 4.7${\times}$ faster than SCA \cite{SCA} and 8.6$\times$ faster than CSCA-ADMM \cite{CSCA-ADMM}.
\hlt{The  gains start to saturate with 2-3 UAVs. In fact,
MAESTRO-X approaches the theoretical lower bound to the delay, for large power consumption values:
with more power available, UAVs leverage their mobility to improve pathloss conditions;
 thanks to spread maximization, multiple UAVs are more likely to be in the vicinity of a request and readily serve it.}

In Fig.~\ref{F9}, we show the contributions of the communication and queue wait times to the overall delay experienced by the GNs, with MAESTRO-X evaluated under a power constraint of 1 kW (less than any other scheme, see Fig.~\ref{F8}).
 We note that the BS-only deployment suffers severely due to large
communication delays of GNs at the cell edge, causing the queue to become backlogged. 
 \hlt{The performance is drastically improved by deploying HAPs (HAP-only), thanks to their 
 higher elevation and improved LoS conditions. Yet, the delay performance offered by a HAP-only deployment is poorer than a non-terrestrial deployment involving UAVs: 2.7$\times$ slower than a static UAV and 3.8$\times$ slower than a UAV optimized with MAESTRO.} 
Across all UAV-assisted implementations, 
 increasing the number of UAVs in the swarm not only lowers the communication delay but also the queue wait times since more GNs can be served simultaneously. \hlt{Remarkably, MAESTRO-X demonstrates negligible queue wait times even with a single UAV: in this low-traffic regime, requests are served quicker than the rate at which they are generated, thereby bypassing the need for piggybacking and frequency reuse.}

\begin{figure} [t]
     \begin{subfigure}{0.55\linewidth}
         \centering
         \includegraphics[width=1.0\linewidth]{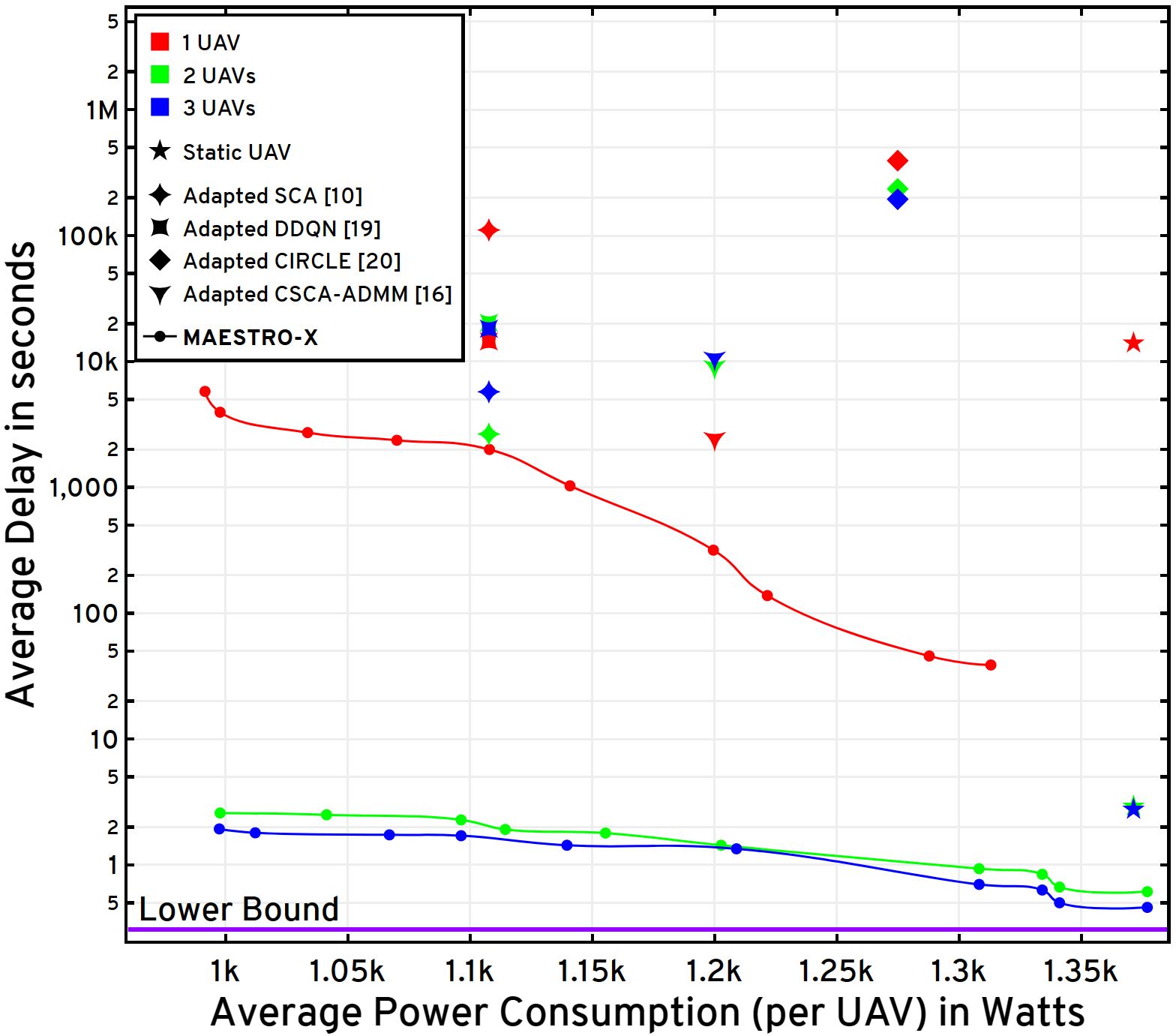}
         \caption{\hlt{Delay-Power Trade-off.}}
         \label{F8a}
     \end{subfigure}
     \begin{subfigure}{0.448\linewidth}
         \centering
         \includegraphics[width=1.0\linewidth]{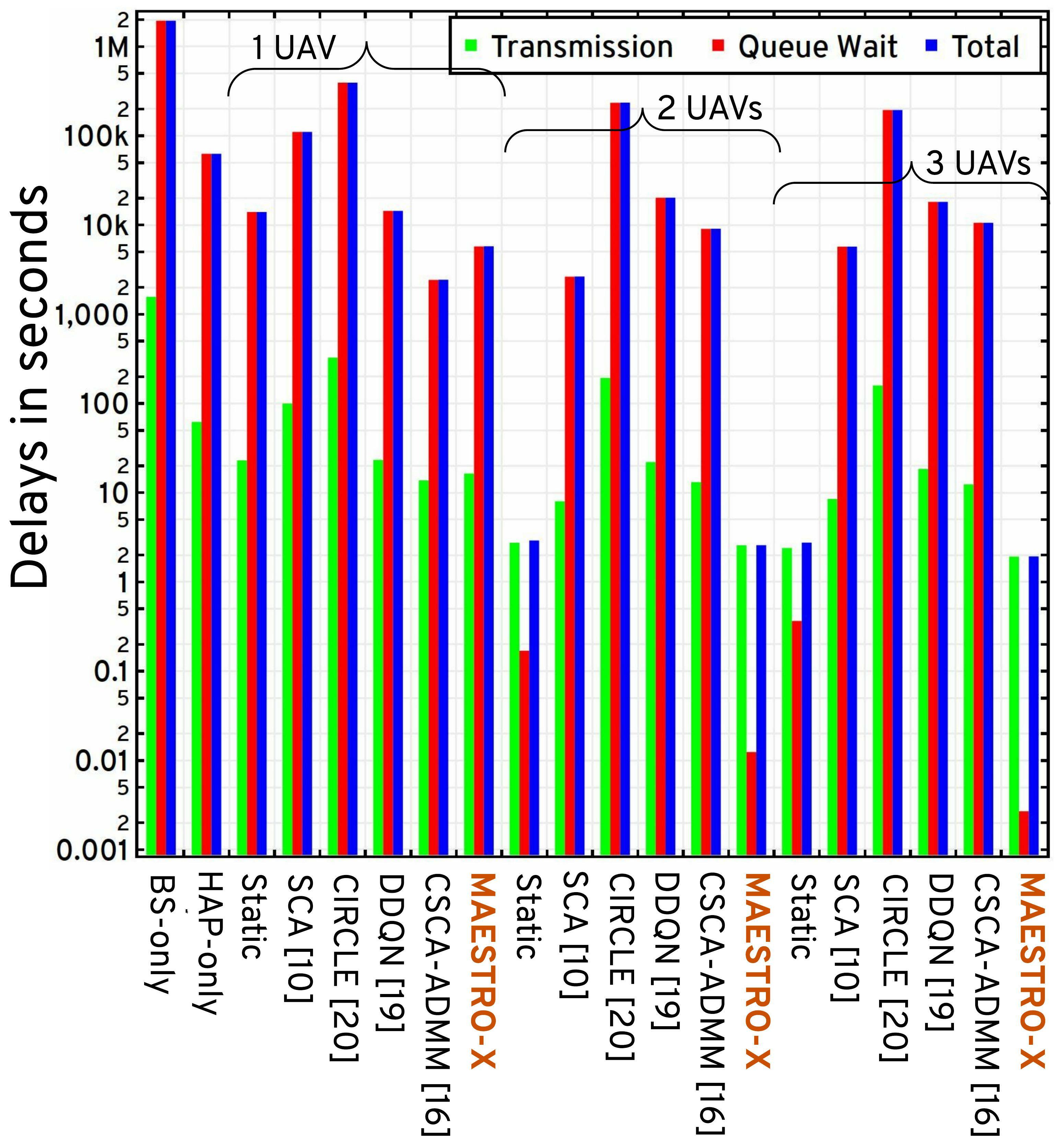}
         \caption{\hlt{Delay chart.}}
         \label{F9a}
     \end{subfigure}
     \caption{\hlt{$L{=}10$ Mbits, $\Lambda'{=}$20 req/min/UAV: Delay-power trade-off (a) and delay charts (b) for MAESTRO-X, state-of-the-art algorithms, and custom heuristics. In (b), MAESTRO-X is evaluated under $P_{\mathrm{avg}}=$1 kW.}}
     \label{F8aF9a}
\end{figure}

\hlt{To analyze the impact of these mechanisms, in Fig.~\ref{F8a} and Fig.~\ref{F9a}, we study a high congestion regime ($\Lambda'{=}$20 req/min/UAV).
The results depicted in Fig.~\ref{F8a} are qualitatively similar to the low congestion case with some key differences: for all the competitor schemes, we note a performance degradation, due to the large wait times (Fig.~\ref{F9a}); a similar performance degradation is noted
for MAESTRO-X with a single UAV.
However, remarkably, MAESTRO-X with  2-3 UAVs  appears to be unaffected by the higher arrival rate, as also demonstrated by the small queue time. This is attributed to 
 frequency reuse allowing more efficient spectrum use, and to piggybacking  allowing simultaneous service of multiple requests by each UAV.}

\begin{figure*}[t]
	\begin{minipage}[b]{0.47\linewidth}
	\centering
	\includegraphics[width=0.95\linewidth]{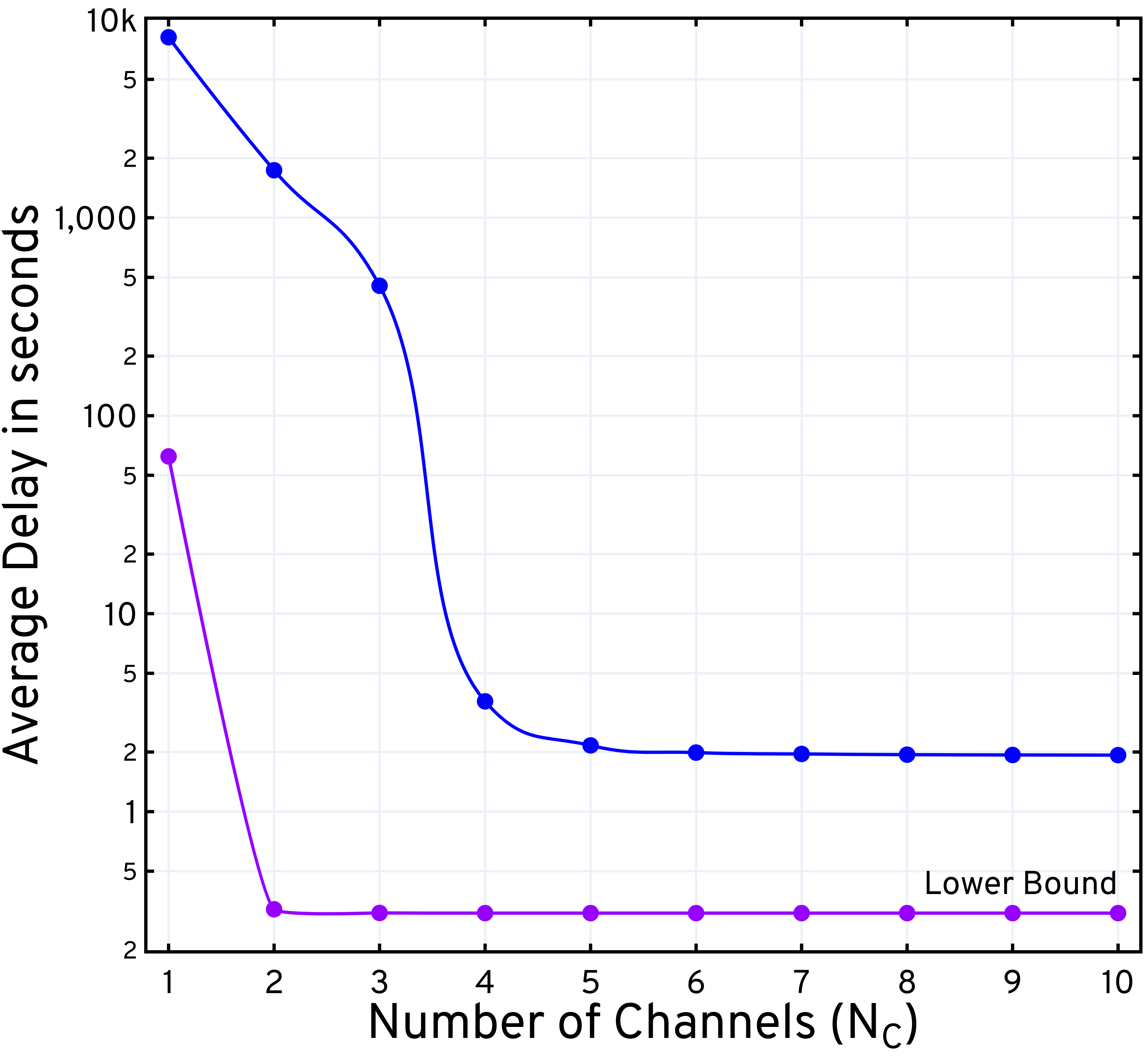}
    \vspace{-2mm}
	\caption{\hlt{10 UAVs, $L{=}$10 Mbits, $P_{\mathrm{avg}}{=}$1 kW, $\Lambda{=}$200 req/min: Average service delay (communication time + queue wait time) vs the number of channels $N_{C}$.}}
	\label{F10a}
\end{minipage}
\hfill
\begin{minipage}[b]{0.485\linewidth}
	\centering
	\includegraphics[width=1.0\linewidth]{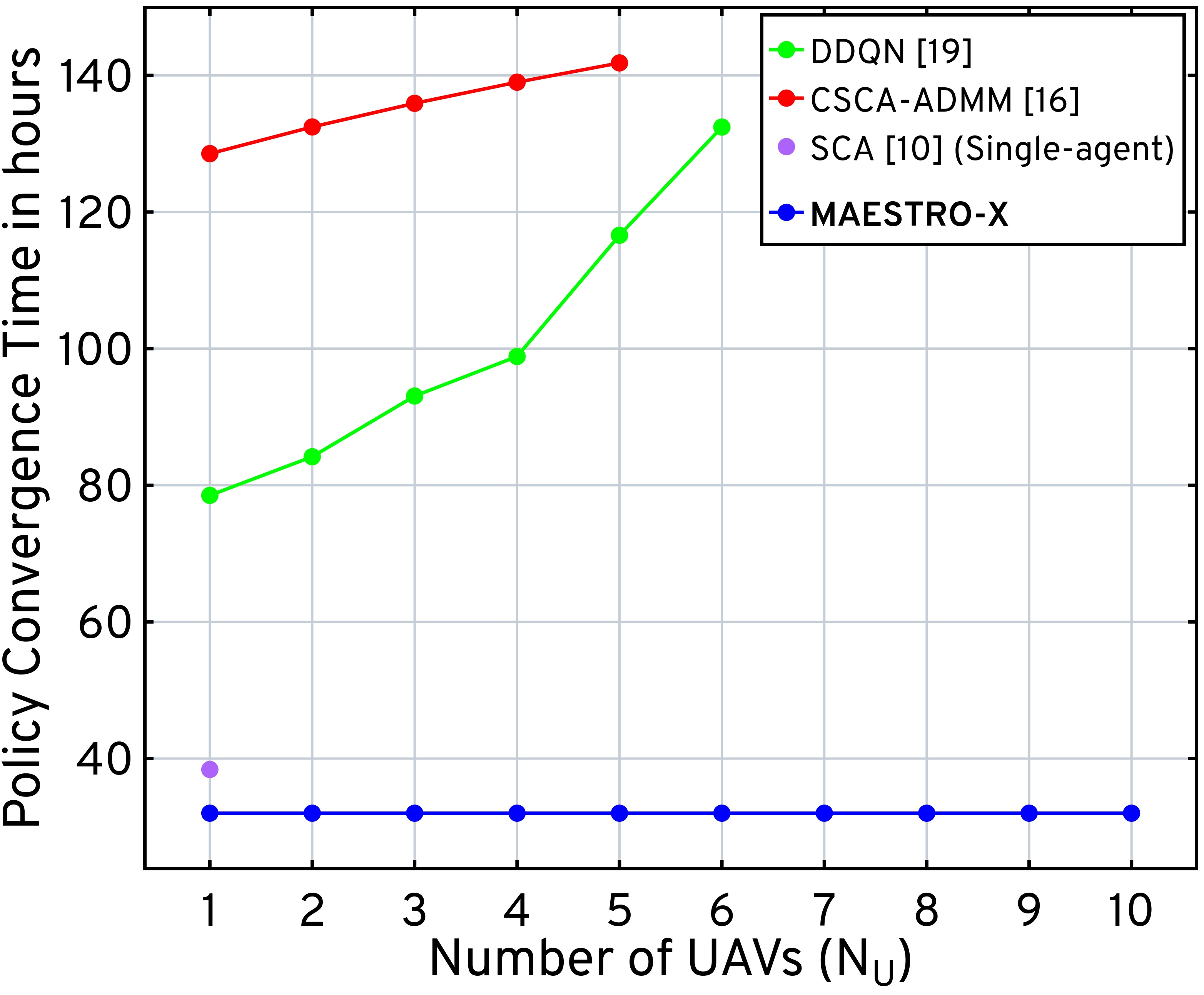}
    \vspace{-2mm}
	\caption{\hlt{Policy convergence time (in hours) for MAESTRO-X and the relevant state-of-the-art.}}
	\label{F10}
\end{minipage}
\end{figure*}

\noindent
\hlt{{\underline{MAESTRO-X, impact of number of channels for large swarms}}:  In Fig.~\ref{F10a}, we study the impact of the number of channels (each of 5 MHz) on the average service delay offered by a MAESTRO-X deployment of 10 UAV-relays, in the high congestion regime.
Note that the competitors become computationally intractable with more than 5-6 UAVs,
whereas the policy replication mechanism of MAESTRO-X offers scalability to large UAV swarms
(see Fig.~\ref{F10}). The delay quickly improves by increasing the number of channels, and saturates after 5 channels at 2s delay (consistent with Fig. \ref{F8a}). This is a remarkable result: for instance, with 4 channels (service delay of $\approx$4 s), if no frequency reuse was allowed, the network could at most service 4[data channels]$\times 15$[req/min/data channel]=60 req/min. The ability to serve a much larger rate of $\Lambda=200$ req/min is attributed to the frequency reuse mechanism.}

\noindent{\hlt{\underline{Policy convergence time}}}: \hlt{Finally, in Fig.~\ref{F10}, we benchmark MAESTRO-X against SCA from \cite{SCA} (single-agent, model-based), CSCA-ADMM from \cite{CSCA-ADMM} (model-based), and DDQN from \cite{DDQN} (model-free), in terms of their policy convergence times, when varying the number of UAVs $N_{U}$. All implementations are in Python, and are executed on a compute node with 2$\times$ 64-core AMD EPYC Milan 7763 CPUs, 16$\times$ 64 GB DDR4 memory, and 4$\times$ NVIDIA A100 GPUs with 40 GB VRAM each. Remarkably, the convergence time of MAESTRO-X is irrespective of the number of UAVs, whereas it grows quickly for CSCA-ADMM and DDQN. This is due to the policy replication mechanism used by MAESTRO-X: the policy is computed for a single-agent, and then replicated across the swarm, coupled with the supplementary UAV-swarm mechanisms developed in Sec.~\ref{S5}. On the other hand, the convergence times of CSCA-ADMM and DDQN grow quickly with the number of UAVs, and become prohibitive when scaled to more than 5 and 6 UAVs, respectively: in fact, it grows linearly for CSCA-ADMM, due to a joint multi-UAV construction involved in its CVXPY-SCS implementation, and exponentially for DDQN, due to combined multi-agent state and action space construction. Remarkably, MAESTRO-X yields a faster convergence time even for a single UAV, thanks to its ability to leverage the multiscale structure of the decision process to achieve a more efficient implementation, in addition to \emph{Tensor}-ized executions exploiting SIMD processing in CUDA-capable GPUs, and distributed workers and thread-pool concurrency in Python (TensorFlow). These benefits in policy convergence coupled with the superior delay-power performance illustrated in Figs.~\ref{F8F9} and \ref{F8aF9a}, present MAESTRO-X as an appealing solution for both small and large UAV swarms.}
\vspace{-4mm}

\section{Conclusion}\label{S7}
In this paper, we propose the MAESTRO-X framework for the decentralized orchestration of rotary-wing UAV-relay swarms in cellular networks, augmenting the coverage and service capabilities of a terrestrial BS. First, we specialize our system to single-UAV deployments and design the optimal scheduling and trajectory optimization policy under an SMDP formulation. Next, we extend to distributed multi-UAV deployments by employing multi-agent coordination mechanisms, and then replicate this augmented single-UAV policy across the swarm. Numerical evaluations demonstrate that MAESTRO-X delivers significant gains over BS- and HAP-only deployments; furthermore, it exhibits superior performance over static UAV deployments, deep Q-learning schemes, and successive convex approximation strategies.
\vspace{-4mm}

\begin{appendices}\label{S8}
\section*{Appendix A: Proof of Prop.~\ref{P1}}\label{X1}
\hlt{Since $2(K{+}1)|g|^{2}$ has a non-central $\chi^{2}$ distribution with $2$ degrees of freedom and a non-centrality parameter $2K$, we find that $P_{\mathrm{out}}(\Upsilon,\beta,K){=}1{-}Q_{1}(\sqrt{2K},\sqrt{2(K{+}1)u(\Upsilon,\beta)}),$ where $Q_{1}(\cdot,\cdot)$ is the standard Marcum $Q$-function \cite{Rician}. Hence,
$
	R(\Upsilon,\beta,K) = \Upsilon\cdot Q_{1}(\sqrt{2K},\sqrt{2(K + 1)u(\Upsilon,\beta)}).
$
We now maximize it over  $\Upsilon{\geq}0$. Let $Z{\triangleq}2\gamma^{-1}u\left(\Upsilon,\beta\right)$ and $\gamma{\triangleq}\frac{N_{0}B}{\beta P}$, hence $\Upsilon{=}B\log_{2}\left(1{+}\frac{Z}{2}\right){\triangleq}f(Z)$. It follows that $\Upsilon^{*}{=}f\left(Z^{*}\right)$, where $Z^{*}$ maximizes over $Z{\geq}0$ the function
\begin{align}\label{eq:Pout_defn}
    h(Z) \triangleq \ln R(f(Z),\beta,K) = \ln f(Z) + \ln Q_1 (\sqrt{2K}, \sqrt{\gamma(K + 1)Z}).
\end{align}
Note that $Q_{1}\left(a,\sqrt{bZ}\right)$ is log-concave in $Z{\geq}0$ for $a,b{>}0$ (see \cite{MarcumTB}), and second derivative of $\ln f(Z)$ satisfies $(\ln f(Z))^{\prime\prime}{=}\frac{f^{\prime\prime}(Z)}{f(Z)}{-}\frac{(f^{\prime}(Z))^{2}}{(f(Z))^2}{\leq}0,{\forall}Z{\geq}0$, so that $h(Z)$ is concave in $Z{\geq}0$. Since $\lim_{Z{\to}0^{+}}h(Z){=}-\infty$ and $\lim_{Z{\to}\infty}h(Z){=}-\infty$, there exists a unique $Z^{*}{\in}(0,\infty)$ (hence $\Upsilon^{*}{=}f\left(Z^{*}\right)$) such that $h^{\prime}(Z^{*}){=}0$, solvable with the bisection method, with $h^{\prime}(Z)$ given by
\begin{align}
    \nonumber
    h^\prime(Z) &= \frac{f^\prime(Z)}{f(Z)} + \frac{\sqrt{\gamma(K + 1)}}{2\sqrt{Z}}\frac{\partial Q_1(\sqrt{2K},b)/\partial b\big|_{b=\sqrt{\gamma(K + 1)Z}}}{Q_1(\sqrt{2K},\sqrt{\gamma(K + 1)Z})},
\end{align}
yielding \eqref{hprime} after solving for $f^{\prime}$ and the partial derivative of $Q_{1}$.}
\vspace{-4mm}

\section*{Appendix B: Proof of Prop.~\ref{P2}}\label{X2}
\hlt{Let $\bar{W}_{\mu}{\triangleq}\bar{W}_{\mu}^{(s)}{+}\bar{W}_{\mu}^{(bs)}$. If $\xi_{u}{=}1$, then additional requests received during the UAV relay phase are served directly by the BS, with delay $\frac{L}{\bar{R}_{GB}(r)}$ for a GN in position $(r,\theta)$. Thus, the expected average communication delay to serve these additional requests is $\mathbb{E}[\Delta_{u,i}^{(bs)}]{=}\bar{D}_{BS}$, yielding $\bar{W}_{\mu}{=}\bar{W}_{\mu}^{(s)}{+}\bar{D}_{BS}(\bar{N}_{\mu}{-}1)$ and $\bar{D}_{\mu}{=}\frac{\bar{W}_{\mu}}{\bar{N}_{\mu}}{=}\frac{\bar{W}_{\mu}^{(s)}}{\bar{N}_{\mu}}{+}\left(1{-}\frac{1}{\bar N_{\mu}}\right)\bar{D}_{BS}$. Let $\mu$ be any policy (including the optimal one) that satisfies $\bar{D}_{\mu}{\leq}\bar{D}_{BS}$: under such policy, since $\bar{N}_{\mu}{\geq}1$, the expression above implies that $\bar{W}_{\mu}^{(s)}{\leq}\bar{D}_{\mu}{\leq}\bar{D}_{BS}$. Moreover, since $\mathbb{E}[N_{u}|\Delta_{u}^{(s)}]{=}\Delta_{u}^{(s)}\Lambda'$ and $\xi_{u}{\leq}1$, it follows that $\bar{N}_{\mu}{\leq}1{+}\Lambda'\bar{W}_{\mu}^{(s)}$ with equality if the UAV always serves requests. This implies \eqref{bounds}.}
\vspace{-4mm}

\section*{Appendix C: Proof of Prop.~\ref{P3}}\label{X3}
\hlt{Let $\pi_{\mathrm{wait}}{=}1{-}\pi_{\mathrm{comm}}$ be the SMDP steady-state probability of the UAV being in the waiting state. Since the probability of remaining in the waiting state (no request is received) in one SMDP step is $p_{ww}{=}e^{-\Lambda'\Delta_{0}}$ and that of moving from a communication state to a waiting state is $p_{cw}{=}1$, $\pi_{\mathrm{comm}}$ and $\pi_{\mathrm{wait}}$ are solutions of the stationary equation
$\pi_{\mathrm{wait}} = \pi_{\mathrm{wait}}p_{ww} + \pi_{\mathrm{comm}}p_{cw}= e^{-\Lambda' \Delta_0}\pi_{\mathrm{wait}} + \pi_{\mathrm{comm}}$.
Solving it with $\pi_{\mathrm{wait}}{+}\pi_{\mathrm{comm}}{=}1$ yields the expression of $\pi_{\mathrm{comm}}$ in Prop.~\ref{P3}.}
\end{appendices}
\vspace{-12mm}

\bibliographystyle{IEEEtran}
\bibliography{IEEEabrv,main}
\end{document}